\documentclass[preprint,preprintnumbers,amsmath,amssymb,apsrev,aps,prb]{revtex4}

\usepackage{graphicx}
\usepackage{dcolumn}
\usepackage{bm}
\begin{document}

\bibliographystyle{plain}
\title{RELATIVE PERIODIC ORBITS\\ IN TRANSITIONAL PIPE FLOW}

\author{Yohann Duguet\footnote{Present address : Linn\'e Flow Centre, KTH Mechanics, SE-100 44 Stockholm, Sweden.}}
\author{Chris C.T. Pringle}
\author{Rich R. Kerswell}
\affiliation{%
School of Mathematics, University of Bristol\\
BS8 1TW, Bristol, United Kingdom}

\begin{abstract}

\small

A dynamical system description of the transition process in shear
flows with no linear instability starts with a knowledge of exact
coherent solutions, among them travelling waves (TWs) and relative
periodic orbits (RPOs). We describe a numerical method to find such
solutions in pipe flow and apply it in the vicinity of a Hopf
bifurcation from a TW which looks to be especially relevant for
transition. The dominant structural feature of the RPO solution is
the presence of weakly modulated streaks. This RPO, like the TW from
which it bifurcates, sits on the laminar-turbulent boundary
separating initial conditions which lead to turbulence from those
which immediately relaminarise.

\end{abstract}

\pacs{Valid PACS appear here}
\maketitle
\newpage

\section{INTRODUCTION \label{secintro}}

Ever since the experiments of Osborne Reynolds in
1883,\cite{Reynolds} it has been known that the flow of a Newtonian
fluid in a circular pipe undergoes an abrupt transition from a
laminar to a turbulent state. This transition is governed by one
dimensionless parameter, the Reynolds number $Re:=Ud/\nu$ ($U$ is
the bulk velocity, $d$ the pipe diameter and $\nu$ the kinematic
viscosity of the fluid). Transition is usually observed
experimentally for $Re \sim 2000 \pm 250$ but can be delayed to
larger values in especially well-controlled
experiments.\cite{Mullin,Pfenniger} The laminar state
(Hagen-Poiseuille flow) is believed linearly stable for all $Re$, so that
transition to turbulence cannot be explained by an ordered sequence
of bifurcations starting from the basic state. Instead, pipe flow
belongs to a more general class of parallel shear flows, including
plane Couette and plane Poiseuille flow, which undergo
finite-amplitude instability.\cite{Drazin} The transitional dynamics
of these flows then appears organised around the presence of other
solutions disconnected from the laminar state.\cite{Waleffe1,Bristol,Kerswell_nonlinearity, Gibson} Such new solutions
are expected to have a relatively simple structure compared to the
turbulent dynamics and to be unstable. These expectations are conceptually satisfying for two reasons. Firstly,
relatively simple three-dimensional mechanisms (e.g. the
`self-sustaining process') have been identified in low-$Re$
flows\cite{Hamilton,Waleffe} allowing turbulence to be sustained
over long times, which evokes the possibility of organised dynamics.
Secondly, instability is needed to rationalise why neither steady
nor periodic exact states are ever found as limiting states to emerge from
an initially turbulent flow. The fact that statistically
recurrent states have been clearly observed, e.g. streaky velocity
fields with an intrinsic wavelength in most
wall-bounded shear flows,\cite{Hamilton} is most easily explained by  
imagining repeated visits in phase space to the vicinity of exact 
solutions of the governing equations.

Progress in numerical computation has first led to the discovery of
some exact solutions in plane shear flows
\cite{Nagata,Waleffe1,Waleffe} and more recently pipe
flow.\cite{Faisst,Wedin} They are all equilibria (steady states) or
relative equilibria (travelling waves which are steady in a moving
frame) and all possess a small number of unstable eigendirections.
They appear through saddle-node bifurcations and are disconnected
from the laminar state. There is a tremendous interest in these
solutions as very similar structures have been observed transiently
in experiments.\cite{Hof, Hof2} It is worth emphasizing that their
simple dynamics (steadiness or constant speed propagation) is
inherent to the method used to seek them, and undoubtedly hides a
larger variety of other solutions. At some point the physical
description of the flow in terms of three-dimensional structures
ceases to be enlightening and it is useful to think in terms of a
trajectory in phase space. Formally, the Navier-Stokes equations can
be projected onto any infinite-dimensional basis of incompressible
velocity fields to yield an autonomous dynamical system of the form
$d{\bm X}/dt={\bm F}({\bm X},Re)$. The choice of the basis, and
hence the particular projection, is not crucial. The topology of the
phase space associated to pipe flow near transition is thought to be
organised around one locally attractive point (the `laminar' flow)
and a chaotic saddle, i.e. a set of unstable solutions and their
heteroclinic and homoclinic connections.\cite{Bristol, Faisst} From
this point of view, it is necessary to know which states the saddle
is made of in order to understand the dynamics near transition. It
has been demonstrated that the unstable solutions known so far in
pipe flow, the travelling waves (TWs), are visited only for about
$O(10$-$20\%)$ of the time.\cite{Tutty,Schneider3} This indicates
that, provided the picture is correct, more solutions of a different
type have to be sought. The next level up in the hierarchy of
solutions from relative equilibria is relative periodic orbits
(RPOs). They are an extension of periodic orbits (in the same way as
traveling waves correspond to degenerate fixed points) due to the
invariance of the equations with respect to the azimuthal and axial
directions. Physically, a RPO corresponds to a flow which repeats
itself periodically in a given moving frame, translating and/or
rotating along the pipe axis at a constant rate. In the cylindrical
coordinate system $(s,\theta,z)$ aligned with the pipe axis, it is a
time-dependent velocity field ${\bm v}(s,\theta,z,t)$ satisfying
\begin{eqnarray}
{\bm v}(s,\theta,z,t+T)={\bm v}(s,\theta+\Delta \theta,z+ \Delta z,t)
\end{eqnarray}
for some constants $T$, $\Delta \theta$ and $\Delta z$.  The initial
motivation behind the search for periodic behaviour is Periodic
Orbit Theory,\cite{ChaosBookRef} which states that any dynamical
average of a smooth chaotic dynamical system can be evaluated in a
deterministic way by finding all unstable periodic orbits of the
system up to a certain period, and carefully averaging over them.
Consideration of continuous symmetries in the original PDEs - here
translation in $z$ and rotation in $\theta$ - extends this expansion
over all \emph{relative} periodic orbits.\cite{Lopez} Even if
complete knowledge of \emph{all} periodic orbits of the system seems
computationally ambitious today, a start needs to be made in
developing numerical tools to find them in anticipation of greater
computing power tomorrow. Some progress has already been made in
plane Couette flow with the recent discovery of periodic\cite{Kida}
and relative periodic orbits\cite{Viswanath} embedded in turbulence,
albeit in small flow domains. Beyond the obvious computational
savings, restricting the largest wavelength in the flow has proved a
very useful tool for isolating key dynamics \cite{Hamilton}.
In the same spirit, we will here concentrate on relatively short
`periodic' pipes only a few diameters long.

In this study we are interested in RPOs which are located in the
laminar-turbulent boundary, the invariant subset of phase-space on
which trajectories neither relaminarise nor become turbulent. States
belonging to this subset can be visited transiently by phase-space
trajectories during the transition process. Phase-space trajectories
constrained to remain precisely on this laminar-turbulent boundary
have been found to approach a chaotic attractor \cite{Schneider2}
centered on one particular TW solution, the `asymmetric TW' found by
Pringle \& Kerswell.\cite{Pringle} The importance of this particular
TW has been confirmed by further calculations which show that
trajectories constrained to lie on this laminar-turbulent boundary
recurrently approach this TW. \cite{Duguet} We therefore concentrate
on looking for RPOs which bifurcate from this asymmetric TW.\\

The paper is organised as follows. Section \ref{sec:computing}
explains in detail the numerical method chosen to find RPOs from a
starting guess and section III describes the new branch of RPOs
found. Subsection \ref{sec:heteroc} confirms that the RPO is on the
laminar-turbulent boundary  and then explores the likelihood of
connections between the RPO and other TWs. Section \ref{sec:disc}
discusses the relevance of the RPO to the transition process as well
as the numerical limitations of the method.

\section{NUMERICAL PROCEDURE \label{sec:computing}}

\subsection{Governing equations}

We consider the incompressible flow of Newtonian fluid in a
cylindrical pipe and adopt the usual set of cylindrical coordinates
$(s,\theta,z)$ and velocity components ${\bm u}= u \hat{{\bm s}} + v
\hat{{\bm \theta}} + w \hat{{\bm z}}$.  Units of length and flow
speed are taken as the pipe radius and the bulk speed $U$. The computational domain is $(s,\theta,z) \in
[0,1]\times[0,2\pi]\times[0,L]$, where $L=2\pi/\alpha$ is the length
of the pipe. The non-dimensionalised incompressible
three-dimensional Navier-Stokes equations read
\begin{eqnarray}
\frac{\partial {\bm u}}{\partial t} + \left({\bm u}\cdot
{\bm \nabla}\right){\bm u} = -
{\bm \nabla}p + \frac{1}{Re}\nabla^{2}{\bm u},
\end{eqnarray}
\begin{eqnarray}
{\bm \nabla} \cdot {\bm u} = 0,
\end{eqnarray}
where $Re:=Ud/\nu$. The flow is driven by a constant
mass-flux condition
\begin{eqnarray}
{1/\pi}\int_{0}^{1}\int_{0}^{2\pi}w \, sdsd\theta=1,
\end{eqnarray}
as in recent experiments.\cite{Mullin, Peixinho, Peixinho2} Although
time is calculated in units of $d/2U$, all times are quoted
hereafter in the usual units of $d/U$. The boundary conditions are
periodicity across the pipe length ${\bm u}(s,\theta,z)={\bm
u}(s,\theta,z+L)$ and no-slip on the walls ${\bm u}(1,\theta,z)={\bm
0}$.  In the non-dimensionalisation used, the expression of the
Hagen-Poiseuille flow is ${\bm u}_{HP}=2(1-s^2)\hat{{\bm z}}$.

\subsection{Time-stepping code}

The basic tool for the numerical determination of periodic orbits is
an  accurate time-stepping code. The direct numerical simulation
code written by A.P. Willis\cite{Willis,Willis2,Willis3} was adopted which is
based upon the  toroidal-poloidal representation of the velocity
field\cite{Marques}
\begin{eqnarray}
{\bm u}= {\bm \nabla} \times (\Psi \hat{{\bm z}}) + {\bm \nabla}
\times {\bm \nabla} \times (\Phi \hat{{\bm z}}).
\label{potentials}
\end{eqnarray}
The two scalar potentials $\Phi$ and $\Psi$ are discretised using
high-order finite differences in the radial direction $s$, and with
spectral Fourier expansions in the azimuthal $\theta$ and
axial $z$ directions. At any discrete radial location $s_j$,
$(j=1,...,N)$, each of the potentials (e.g. here $\Phi$) is therefore
discretised according to the formula:
\begin{eqnarray}
\Phi(s_j,\theta,z,t;\alpha)= \sum_{k=-K}^{K}
\sum_{m=-M}^{M}\Phi_{jkm}(t)e^{i(m\theta +\alpha k z)}
\label{representation}
\end{eqnarray}
The resolution of a given calculation is described by a triplet
$(N,M,K)$. The phase-space associated to the Navier-Stokes equations
is the set of complex coefficients $\{\Phi_{jkm},\Psi_{jkm}\}$,
which corresponds to a dynamical system with $n \sim 8MNK$ real
degrees of freedom (typically $n=O(10^{5})$). Its metric is defined
by the Euclidean scalar product $\langle \, , \,  \rangle$. A shift
back in physical space by $(\Delta z, \Delta \theta)$ corresponds in
phase-space to the transformation:
\begin{eqnarray}
(\Psi_{jkm},\Phi_{jkm}) \rightarrow
(\Psi_{jkm},\Phi_{jkm})e^{-i(m\Delta \theta + \alpha k \Delta z)}
\label{formulashifts}
\end{eqnarray}
Time discretisation is of second-order, using a Crank-Nicholson
scheme for the diffusion term and an Euler predictor step for the
non-linear terms. In this study we confine our calculations to a
pipe of length $L = 2\pi /0.75 \sim 8.38 $ radii $=4.19\, d$. This length has been chosen because it is a wavelength for which the asymmetric
TW\cite{Pringle} is  known to be on the laminar-turbulent boundary
and for which most data were already available. \\

\subsection{The Newton-Krylov method \label{Newton}}

\subsubsection{Periodic orbits}

\begin{figure}
\includegraphics[width=10cm]{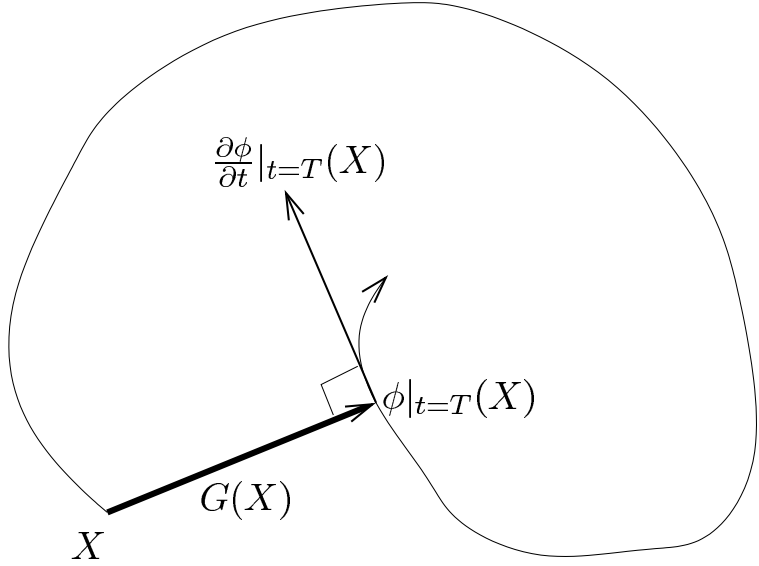}

\caption{Schematic view of phase-space. Definition of the functional
 ${\bm G}$ to minimise in order to find exact periodic orbits.}
\label{UPO}
\end{figure}

A periodic orbit (of period $T$) of a dynamical system $\dot{{\bm
X}}={\bm F}({\bm X})$ is sought as a solution $({\bm X},T)$ of the
equation ${\bm G}({\bm X})={\bm 0}$, where
\begin{eqnarray}
{\bm G}({\bm X}):=\phi^{T}({\bm X})-{\bm X}.
\end{eqnarray}
Here $\phi^{t}({\bm X})$ refers to the point at time $t>0$ on the
trajectory starting from ${\bm X}$ at time $t=0$.  A value of the
period $t=T$ has to be chosen to uniquely define ${\bm G}({\bm X})$.
We choose $T$ as the time minimising the ${\bm X}-$dependent scalar
function $g:t \rightarrow |\phi^{t}({\bm X})-{\bm X}|^{2}$ over a
given time interval. This is done by accurate interpolation of the
zeros of its derivative,
 the scalar function $g':t\rightarrow
 2 \langle \phi^{t}({\bm X})-{\bm X},\partial \phi^{t}({\bm
X})/\partial t \rangle$. Figure \ref{UPO} shows schematically that the time
$t=T$ is picked up along the trajectory when its tangent vector is
orthogonal to the difference vector $\phi^{t}({\bm X})-{\bm X}$,
which ensures that the trajectory has come back closest to the
initial point. A standard method to search for zeros of ${\bm G}$ in
low-dimensional systems (typically $n \lesssim O(10^3)$) is the
Newton-Raphson method. This iterative algorithm, based on successive
Taylor expansions of ${\bm G}$, produces a sequence of iterates
${\bm X}^{(k)},~k \ge 0$ defined by
\begin{eqnarray}
{\bm X}^{(k+1)}={\bm X}^{(k)}+\bm{\delta X}^{(k)},
\end{eqnarray}
where
\begin{eqnarray}
{\bm J}({\bm X}^{(k)})\bm{\delta X}^{(k)}=-{\bm G}({\bm X}^{(k)}),
\label{newtonstep}
\end{eqnarray}
%
%
${\bm J}$  being the Jacobian matrix associated to ${\bm G}$. When no
analytical expression for ${\bm J}$ is available, it is
generally computed by finite differences, at the cost of one
evaluation of ${\bm G}$ for each column of the matrix. In very high
dimension $n \gtrsim O(10^{4})$, however, the time needed for $n+1$
evaluations of ${\bm G}$, as well as the storage of ${\bm J}$
(not to mention solving equation (\ref{newtonstep})), can become
prohibitive, thus we turn our attention towards matrix-free Inexact
Newton-Krylov methods.

At every Newton step $k$, we solve the system (\ref{newtonstep})
using the GMRES algorithm.\cite{Saad} This iterative linear solver
only requires matrix-vector products that can be evaluated using
finite differences, since for a vector ${\bm p}$ and $\epsilon$
sufficiently small (e.g. $10^{-7}$) we have:
\begin{eqnarray}
{\bm J}{\bm p} = \frac{{\bm G}({\bm X}^{(k)}+\epsilon {\bm
p})-{\bm G}({\bm X}^{(k)})}{\epsilon}+O(\epsilon). \label{Jp}
\end{eqnarray}
Hence no matrix needs to be stored explicitly. Using (\ref{Jp}),
GMRES iteratively builds an orthogonal basis of the Krylov subspace
spanned by $\{{\bm r}_0,{\bm J}{\bm r}_0,{\bm J}^{2}{\bm
r}_0,...\}$, starting from ${\bm r}_0=-{\bm G}({\bm X}^{(k)})$. At
each GMRES step, a Gram-Schmidt procedure reorthonormalises the
Krylov subspace producing a basis $\{ {\bm V}_{0},{\bm V}_{1},{\bm
V}_{2},...\}$. Then an approximation to the solution of
(\ref{newtonstep}) is constructed on this basis, until convergence
is decided according to the criterion:
\begin{eqnarray}
|{\bm J}^{(k)}\bm{\delta X}^{(k)} + {\bm G}^{(k)}| \le
 \eta^{(k)} |{\bm G}^{(k)}|.
\label{inexact}
\end{eqnarray}
The `forcing' constant $\eta^{(k)}$ appearing in formula
(\ref{inexact}) is ideally zero, and numerically it should be in
principle very small. However, non-vanishing values of $\eta$ can
allow for a much quicker convergence of the overall Newton scheme, by
avoiding useless oversolving. Values of $\eta \sim O(10^{-1})$ have
produced faster convergence and even better performances were observed
when updating $\eta^{(k)}$ with the choice 2 of Eisenstat \&
Walker.\cite{Walker}

\subsubsection{Double dogleg step}

Newton-Raphson algorithms are known to converge only if the initial
guess is sufficiently close to a zero of the function ${\bm G}$,
which in high dimension often means no convergence at all. Moreover,
even close to a solution, Newton-Raphson steps either can be too
large or nearly orthogonal to the gradient of $|{\bm G}|^2$, in
which case the algorithm stagnates and the classical strategy of
linear backtracking (also called `damped Newton') is of no
help.\cite{Walker2} To remedy this, it is useful to embed the
algorithm into what is commonly called a `globally' convergent
strategy. Here `global' does not mean that convergence is achieved
whatever the starting point, but instead that the basin of
attraction of a solution is reasonably enlarged. We adopt the
`double dogleg step' method proposed by Dennis and
Schnabel,\cite{DennisSchnabel} which is a subclass of the `trust
region' algorithms (see Viswanath\cite{Viswanath} for an implementation of
another subclass call the `hook step' approach).
All these methods start from the knowledge of the Newton step
$\bm{\delta X}_{N}$ (here we omit the
superscript $(k)$\,), and aim at constructing a new vector
$\bm{\delta X}$ whose norm is bounded by a `trust length'
$\delta_c$. The way $\bm {\delta X}$ is chosen is described in the next paragraph.
At the first stage of the loop, $\bm{\delta X}=\bm{\delta X}_{N}$
and $\delta_{c}=|\bm{\delta X}_N|$.
Then at every iteration of the trust region algorithm, a check is made to see
whether $|{\bm G}(\bm{X^{(k)}}+\bm{\delta X})|$ is sufficiently
smaller than  $|{\bm G}(\bm{X}^{(k)})|$. If it is, then the loop
is finished and we set ${\bm X}^{(k+1)}$ to ${\bm X}^{(k)}+\bm{\delta X}$;
if not, then  $\delta_{c}$ is divided by 2, a new vector  $\bm{\delta X}$
is computed and the process is repeated.

In the case of the double dogleg step, two descent directions for
$|{\bm G}|^{2}$ are used to generate the new vector $\bm{\delta X}$
: the first one is the Newton step $\bm{\delta X}_N$, the second is
the steepest descent direction (see figure \ref{dogleg}):
\begin{eqnarray}
-\frac{1}{2}\nabla |{\bm G}|^{2}=-{\bm J}^{t}{\bm G},
\end{eqnarray}
where ${\bm J}^{t}$ is the transpose of the Jacobian matrix
${\bm J}$. We can define the Cauchy Point ${\bm X}^{(k)} +
\bm{\delta X}_{CP}$, where the vector $\bm{\delta X}_{CP}$ minimizes
the quadratic form $\bm{\delta X} \rightarrow |{\bm J}
\bm{\delta X}+{\bm G}|^{2}$ along the steepest descent
direction (whereas the overall minimizer of this quadratic form
defines the Newton step $\bm {\delta X}_N$). Projection of
$\bm{\delta X}_{CP}$ onto the Krylov space (already built to find
$\bm{\delta X}_{N}$ at the same Newton iteration) allows a low-cost
approximation of $\bm{\delta X}_{CP}$, as long as the dimension of
the Krylov subspace itself is low. If the $\bm{V}_{i}$\,s form an
orthonormal basis of the Krylov subspace and $K$ is its dimension,
we have:
\begin{eqnarray}
{\bm \nabla} |{\bm G}|^{2} \sim \sum_{i=1}^{K} \langle{\bm
J}^{t}{\bm G},{\bm V}_{i} \rangle {\bm V}_{i}=\sum_{i=1}^{K}
\langle {\bm G},{\bm J}{\bm V}_{i} \rangle {\bm V}_{i}.
\label{doglegtrick}
\end{eqnarray}
In (\ref{doglegtrick}), the scalar coefficients $\langle {\bm
G},{\bm J}{\bm V}_{i} \rangle$ are already known up for $1\leq
i \leq K$ since (normalized by $-|{\bm r}_0|$) they form the first
row of the
Hessenberg matrix used for the GMRES inversion.\cite{Saad}\\

Finally, given $\bm{ \delta X}_{N}$ and $\bm{ \delta X}_{CP}$, the
double dogleg algorithm finds the intersection between the piecewise
linear curve linking the Cauchy Point to ${\bm X} + \mu \bm{\delta X}_{N}$, and the ball of radius $\delta_{c}$. So the new trial vector is
chosen as:
\begin{eqnarray}
\bm{\delta X} = \bm{\delta X}_{CP}  + \lambda_{c} \left(\mu \bm{\delta X}_{N} -  \bm{\delta X}_{CP} \right).
\label{dogleg}
\end{eqnarray}
where $\mu$ is a constant set to $0.8$ (following ref. \cite{DennisSchnabel})
and $\lambda_{c}>0$ is chosen such that $|\bm{\delta X}|=
\delta_c$. If $|\delta_c| \le |\bm{\delta X}_{CP}|$, the dogleg
curve along which optimisation is achieved is simply the
steepest-descent direction of $|{\bm G}|^{2}$.

\begin{figure}
\includegraphics[width=10cm]{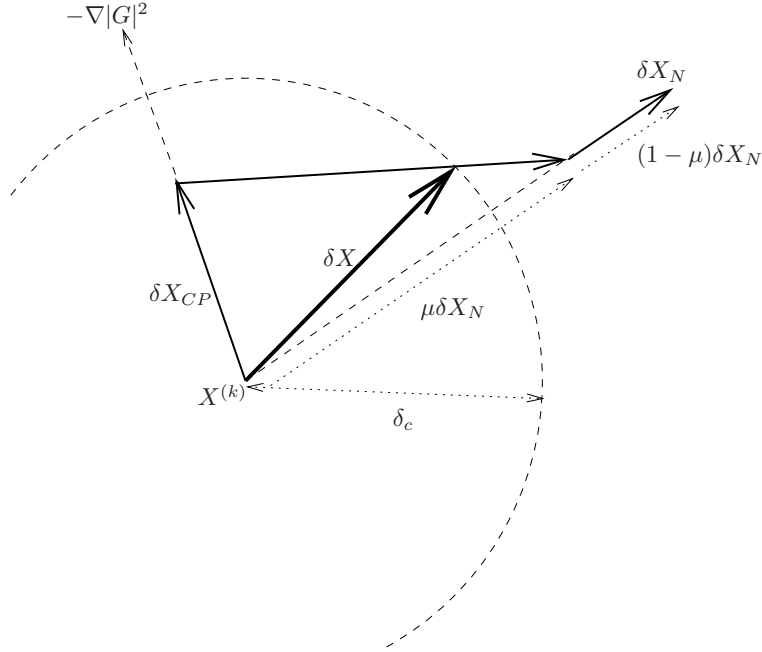}
\caption{Schematic view of the double dogleg technique (adapted from
Dennis \& Schnabel\cite{DennisSchnabel}) in phase space. ${\bm
X}^{(k)}$ is the state at the $k^{th}$ Newton iteration, $\bm{\delta
X}_{N}$ represents the Newton step computed using the inexact GMRES
method, and $-{\bm \nabla}|{\bm G}|^{2}$ is the steepest descent
direction of the residual from ${\bm X}^{(k)}$. At a given step, a new
state ${\bm X}^{(k+1)}=\bm{X^{(k)}+\delta X}$ is suggested as the intersection between the
double dogleg curve (medium thick line) and the ball of radius $\delta_{c}$.}
\label{dogleg}
\end{figure}

\subsubsection{Relative periodic orbits \label{shifts}}

Allowing for shifts in $\theta$ and $z$ implies a modification of the
function $\bm{G}$.  If we seek a relative periodic orbit satisfying
${\bm v}(r,\theta,z,t+T)={\bm v}(r,\theta + \Delta \theta,z + \Delta
z,t)$ for any $t$, we have to change ${\bm G}({\bm X})$ into ${\bm
G}({\bm X})=\phi^{T}_{-\Delta \theta,-\Delta z}({\bm X})-{\bm X}$. In
this expression, $\phi^{T}({\bm X})$ is shifted back in the $\theta$-
and $z$-directions using (\ref{formulashifts}), before being compared
to the initial point of the orbit. $T$ is chosen as the time
minimising the new $|{\bm G}|^{2}$, whereas $\Delta
\theta$ and $\Delta z$ are considered as two extra variables of the
Newton algorithm:\cite{Viswanath} we define a $(n+2)$-dimensional vector
${\bm X}^{+}=({\bm X},\Delta \theta, \Delta z)$ and the function ${\bm
G}^{+}:R^{n+2}\rightarrow R^{n+2}$ such that
\begin{eqnarray}
{\bm G}^{+}_{i}({\bm X}^{+})={\bm G}_{i}({\bm X}),~(i=1,..,n),
\end{eqnarray}
while the two last components of ${\bm G}^{+}$ are defined
by the scalar products in $R^n$:
\begin{eqnarray}
{\bm G}^{+}_{n+1}({\bm X}^{+})=\left \langle {\bm G},\frac{\partial \phi^{T}}{\partial
\Delta \theta} \right \rangle,
\end{eqnarray}
\begin{eqnarray}
{\bm G}^{+}_{n+2}({\bm X}^{+})=\left \langle{\bm G},\frac{\partial \phi^{T}}{\partial
\Delta z} \right \rangle.
\end{eqnarray}

These two derivatives are easily computed by central finite
differences. Starting from a guess ${\bm X}_{0}$ (which includes a
starting point in phase space as well as two initial values of
$\Delta \theta$ and $\Delta z$), the Newton-Krylov method is now
applied to locate zeros of the function ${\bm G}^{+}$. It is worth
reiterating that although $T$ looks to be on the same footing as
$\Delta \theta$ and $\Delta z$, this is not the case. The condition
$< {\bm G},\partial \phi^t/\partial t>=0$ which defines $T$ is
imposed at every Newton/dogleg iteration whereas the equivalent
conditions for $\Delta \theta$ and $\Delta z$ are  only truly
reached when convergence to a RPO has been achieved.

Non-rotating TW solutions are a special case of RPOs, when $T$ and
$\Delta z$ are linked by the relation $\Delta z = cT$, with $c$ the
axial propagation speed of the wave. Because of this degeneracy, TW
solutions with zero azimuthal speed can be sought, using the same
algorithm, by fixing the value of $\Delta z$ to $2\pi/\alpha$, and
setting ${\bm G}^{+}_{n+2}=0$.

\section{RESULTS \label{sec:results}}

\subsection{Hopf bifurcation of traveling waves \label{TW}}

\begin{figure}
\includegraphics[width=8cm]{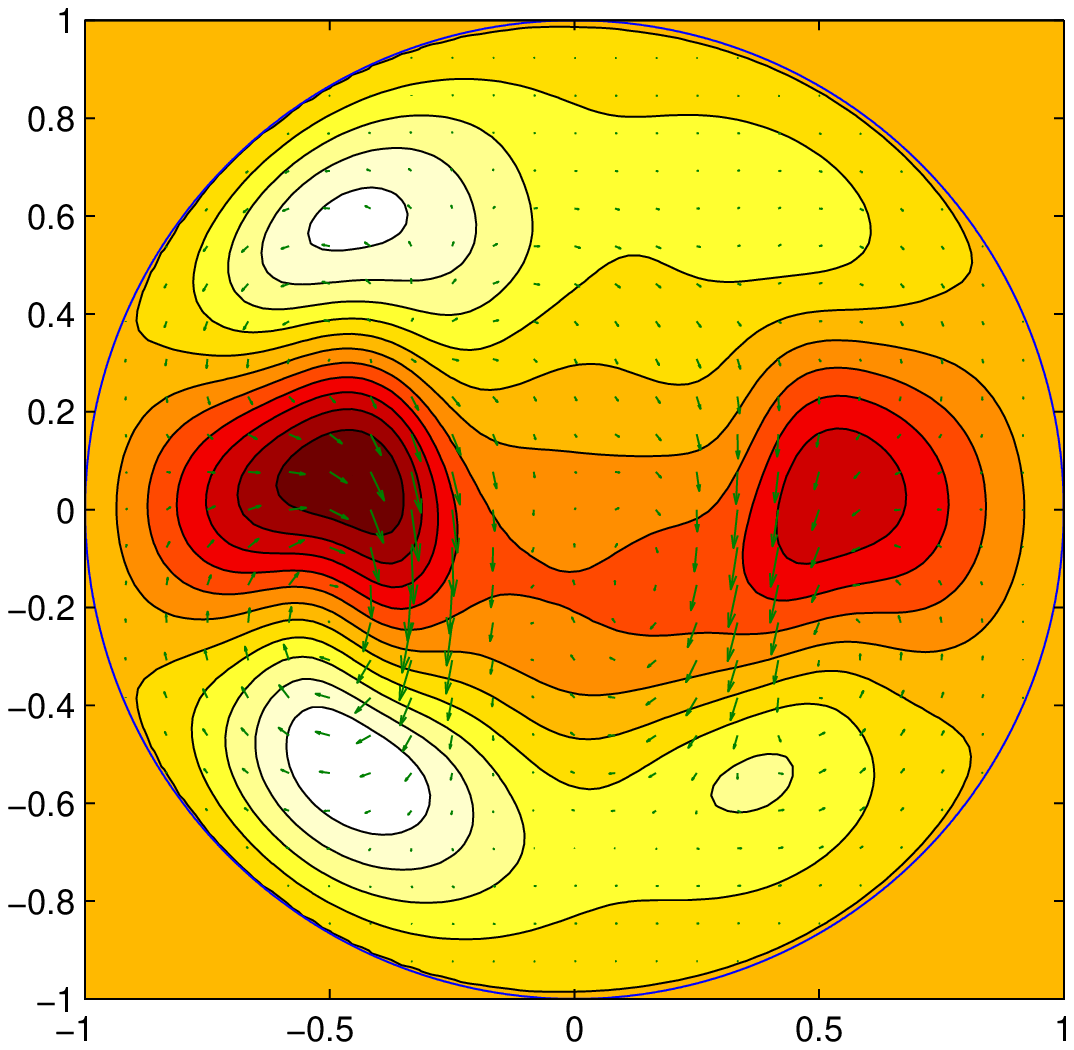}
\includegraphics[width=8cm]{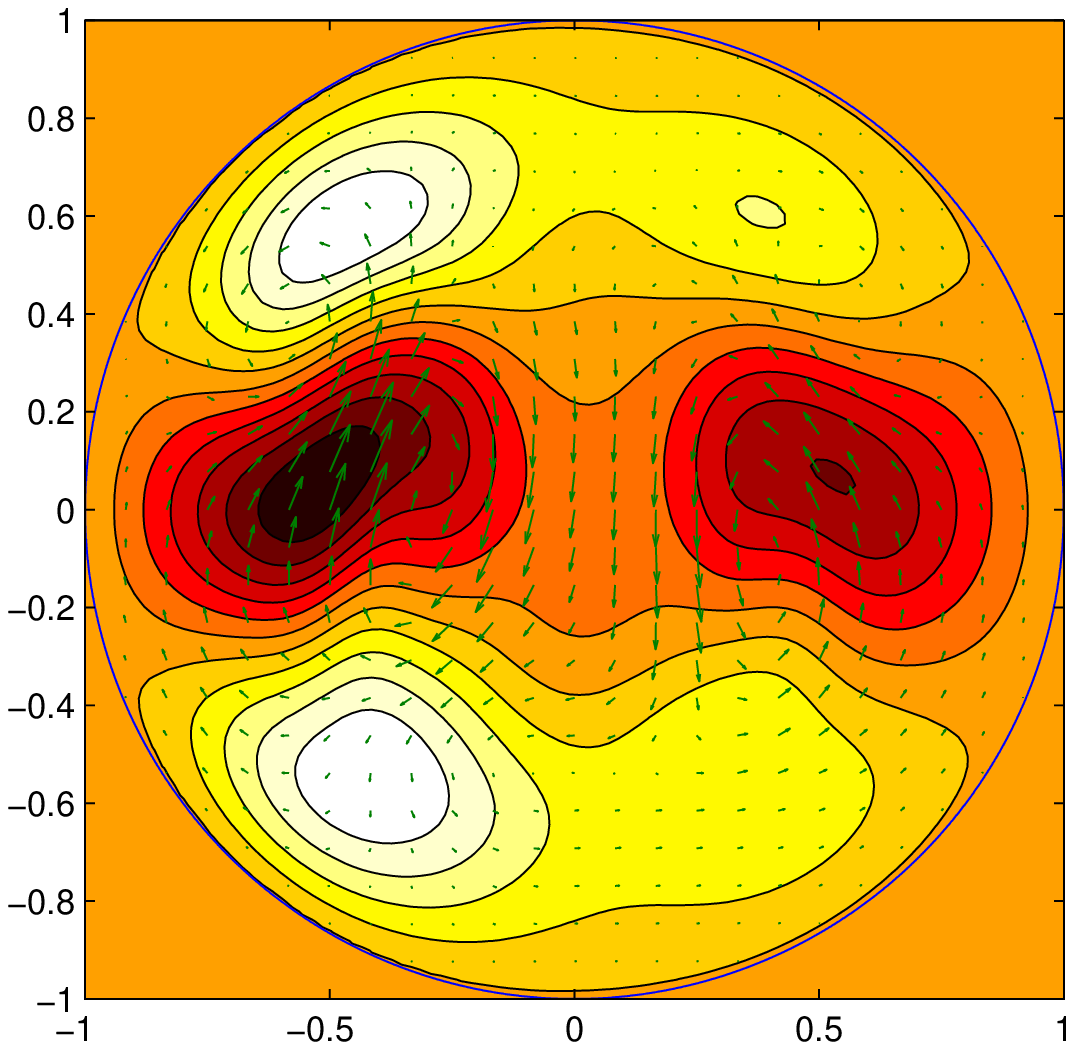}\\
\includegraphics[width=8cm]{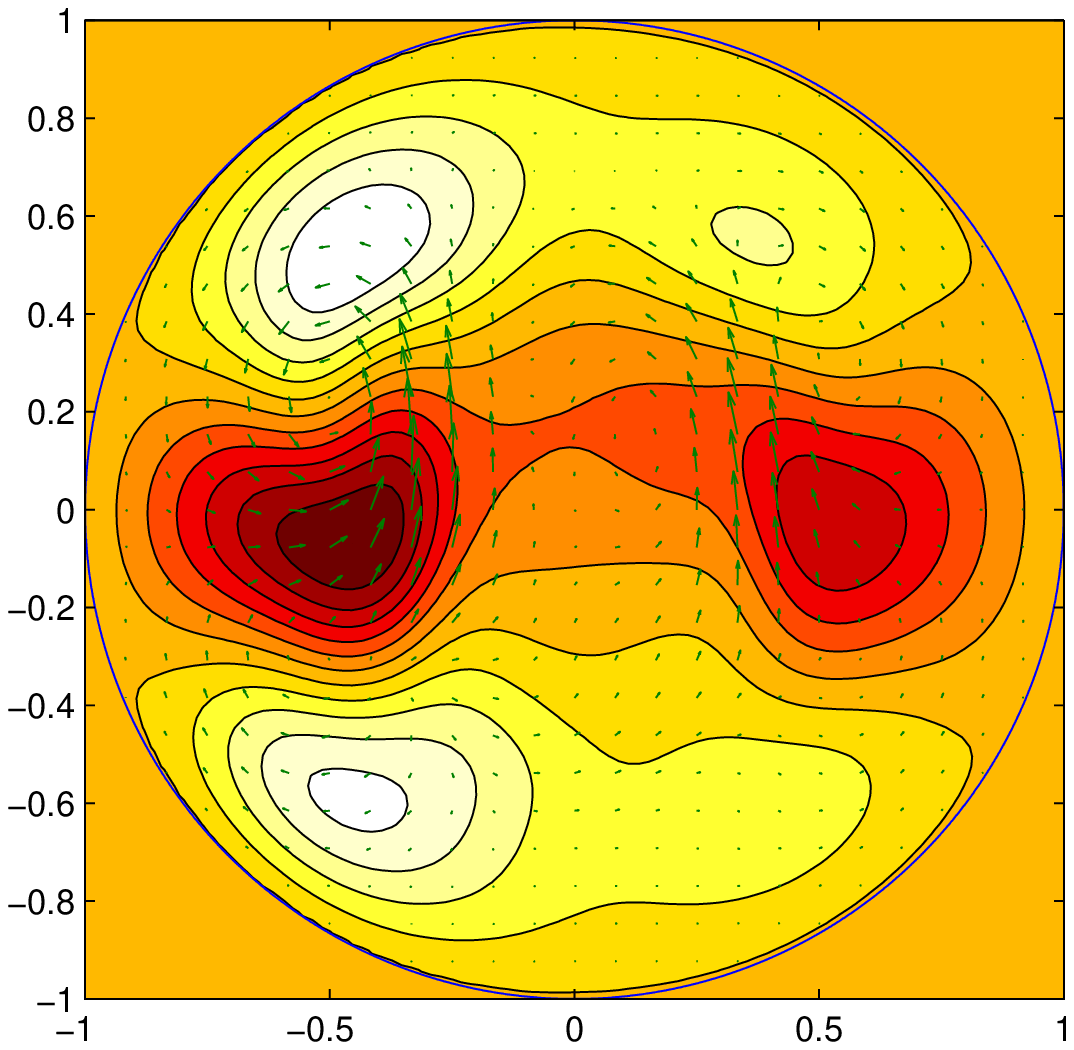}
\includegraphics[width=8cm]{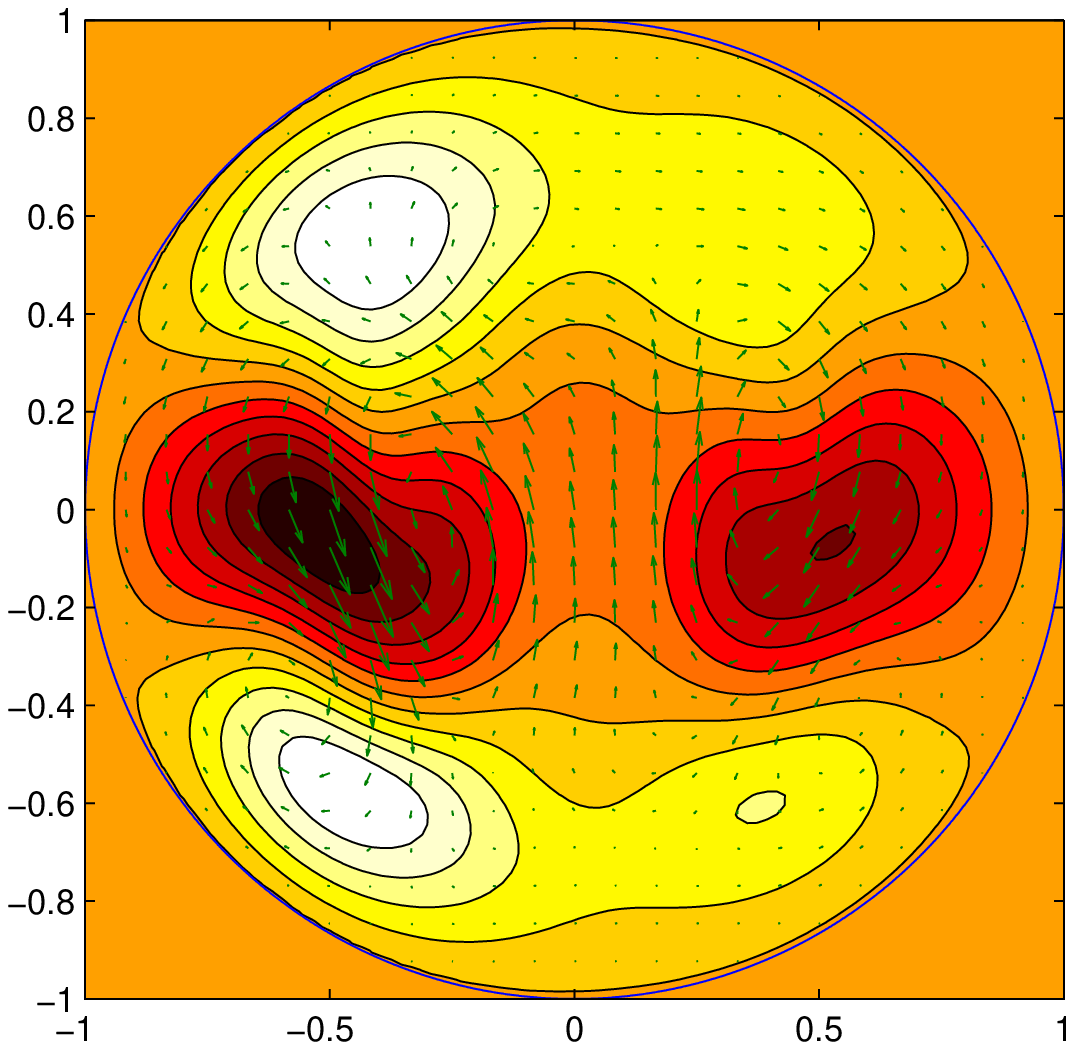}\\
\caption{Velocity profile of the asymmetric TW solution at the Hopf
bifurcation ($\alpha=0.75$, $Re =1785$). From left to right, from top to bottom :
$z/L=0,0.25,0.5,0.75$. The arrows indicate the cross-stream
velocity, and the shading indicates the difference between the
streamwise velocity and the laminar profile (light/white indicating
positive values and  dark/red negative values).} \label{plots0}
\end{figure}

The Newton-Krylov algorithm described in Section \ref{sec:computing}
was used to find a RPO bifurcating off the asymmetric TW branch in a
periodic pipe of length $L \sim 4.19\,d$ ($\alpha=0.75$)
where it is known to be embedded in the laminar-turbulent
boundary. The asymmetric TW branch originates from a
supercritical bifurcation at $Re \sim 1770$ off a
``mirror-symmetric" branch of TWs.\cite{Pringle} The asymmetric TW
propagates axially with a phase speed $c=1.34\, U$ but does not
rotate. For $Re \sim 1785$, it displays a pair of high-speed streaks
sandwiching an eccentric low-speed streak (see Figure \ref{plots0}).
A streamwise vortex tube is present in the vicinity of the low-speed
streak, ensuring the regeneration of the streak via a lift-up
effect. The two high-speed regions extend to the other side of the
cross-section, and another low-speed streak of lesser intensity is
present on the right of the original one. As $Re$ increases, the
solution becomes more asymmetric and the streaks on the right of the
Figure tend to disappear.

\begin{figure}
\includegraphics[width=14cm]{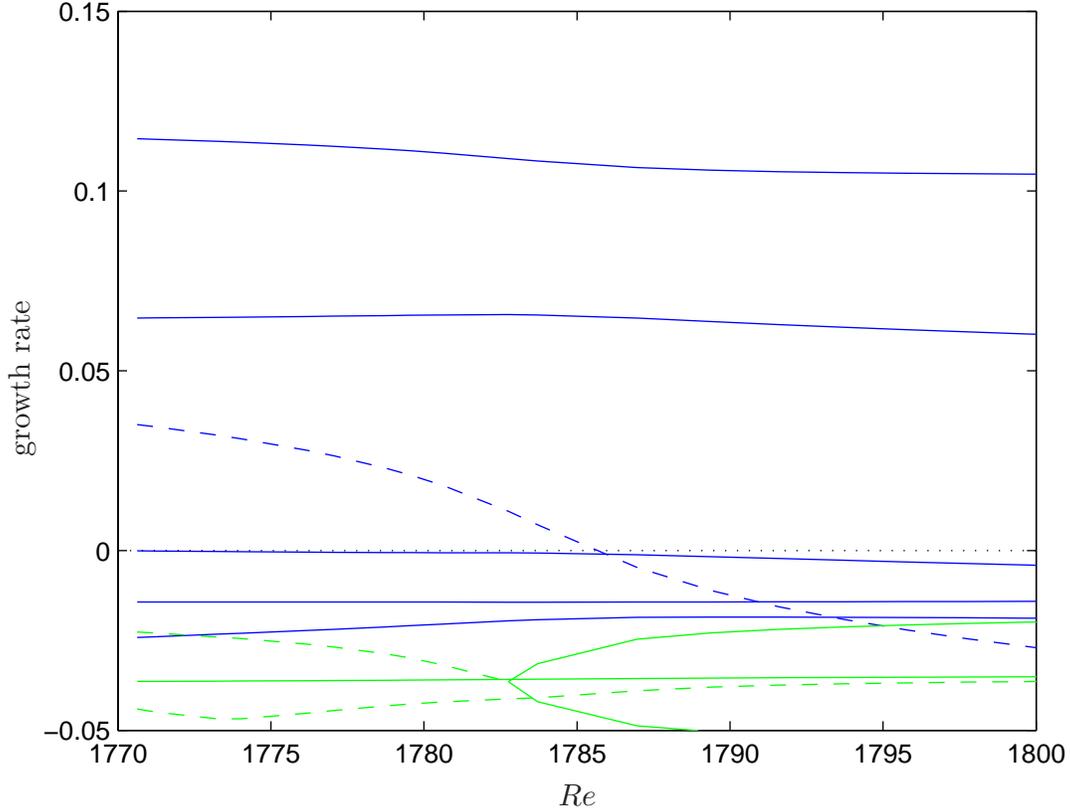}
\caption{Stability analysis of the asymmetric TW for $\alpha=0.75$
near the Hopf bifurcation point at $Re=Re_H=1785.6$: growth rates
(in units of U/D) against $Re$. Each line either indicates the locus
of a real eigenvalue (solid) or a complex conjugate pair (dashed) as
$Re$ changes. The blue (dark) lines correspond to those which are
shift-\&-reflect symmetric\cite{Wedin} while the green (light) lines
are anti-shift-\&-reflect symmetric. The TW itself bifurcates off
the mirror-symmetric TW branch near $Re=1770$.} \label{Eigenvalues}
\end{figure}

The numerical resolution chosen in this work is $(N,M,K)=(86,16,5)$
to match that used by Pringle \& Kerswell in the $\theta$- and
$z$-directions. This is sufficient to observe a drop-off of 6
decades in the axial spectrum and more than 8 decades in the radial
and azimuthal energy spectra. The rapid drop-off in the axial
wavenumbers is typical of lower-branch solutions and their weakly
wiggling streak structure. It is even more pronounced as $Re
\rightarrow \infty$.\cite{Wang} The Navier-Stokes equations have
been linearised around the TW solution expressed in the Galilean
frame moving with speed $c$ along the axis. The corresponding
eigenvalue problem was solved numerically using an Arnoldi routine
for $Re$ up to $5000$.\cite{arpack} There are 4 unstable eigenvalues
(2 real and 2 complex conjugate ones) for $Re<Re_{H} = 1785.6$ and
only 2 real ones for $Re>Re_{H}$ indicating a Hopf bifurcation at
$Re=1785.6$ (see Figure \ref{Eigenvalues}).  The center manifold
theorem states that for $Re$ close enough to $Re_{H}$, a periodic
orbit must exist in the same moving frame. Its amplitude scales
locally like $O(\sqrt{|Re_{H}-Re|})$.\cite{Kuznetsov} The period of
the orbit is given by the imaginary part of the Hopf eigenvalue
pair, here $T = 2\pi/Im(\lambda_H) \sim 43\,d/U$. A good estimation
for the shifts is the distance travelled by the TW in the axial and
azimuthal directions during this period, hence $\Delta z^{(0)} = cT
\sim 58.7\,d$ and $\Delta \theta^{(0)}=0$ (since this wave has no
azimuthal phase speed). The starting point used for the
Newton-Krylov algorithm is a slight perturbation of the TW along
\emph{one} of the two conjugate neutral directions, denoted as the eigenvector ${\bm e}_H$:
 \begin{eqnarray}
    \left({\bm X}^{+}\right)^{(0)}=\left({\bm X}_{TW} + \epsilon {\bm e}_{H},\Delta z^{(0)}, \Delta \theta^{(0)} \right).
    \label{excitation}
 \end{eqnarray}
For values of $\epsilon \sim O(10^{-2})$ and $Re_{H}-1<Re<Re_{H}$,
the algorithm converged to a RPO, within a residual
$r_{\infty}=|{\bm X}_{-\Delta z}(T)-{\bm X}(0)|/|{\bm X}(0)|$ of
$O(10^{-6})$. The converged values of the period $T$ and the shift
$\Delta z$ are very close to the expected values, while the RPO
appears not to rotate ($\Delta \theta=0$). 
The new solution branch bifurcates towards the regime where the original branch has fewer stable directions, namely $Re<Re_{H}$. This Hopf bifurcation is hence supercritical, although the new solution branch extends to lower values of $Re$. From now on we will also use the reduced positive parameter $\delta = Re_{H}-Re$.

\subsection{Continuation of the RPO}

Continuation of these orbits along the $Re$-axis is used to produce
a bifurcation diagram.  Once one RPO is known at a given value of
$\delta$, it is straightforward to use it as an initial condition
for the Newton-Krylov algorithm at slightly different values of
$\delta$. Progression towards larger $\delta$ (i.e. lower $Re$) by
using small enough steps enables one in principle to track the
solution down until the branch
possibly bends back. \\

\begin{figure}
\includegraphics[width=14cm]{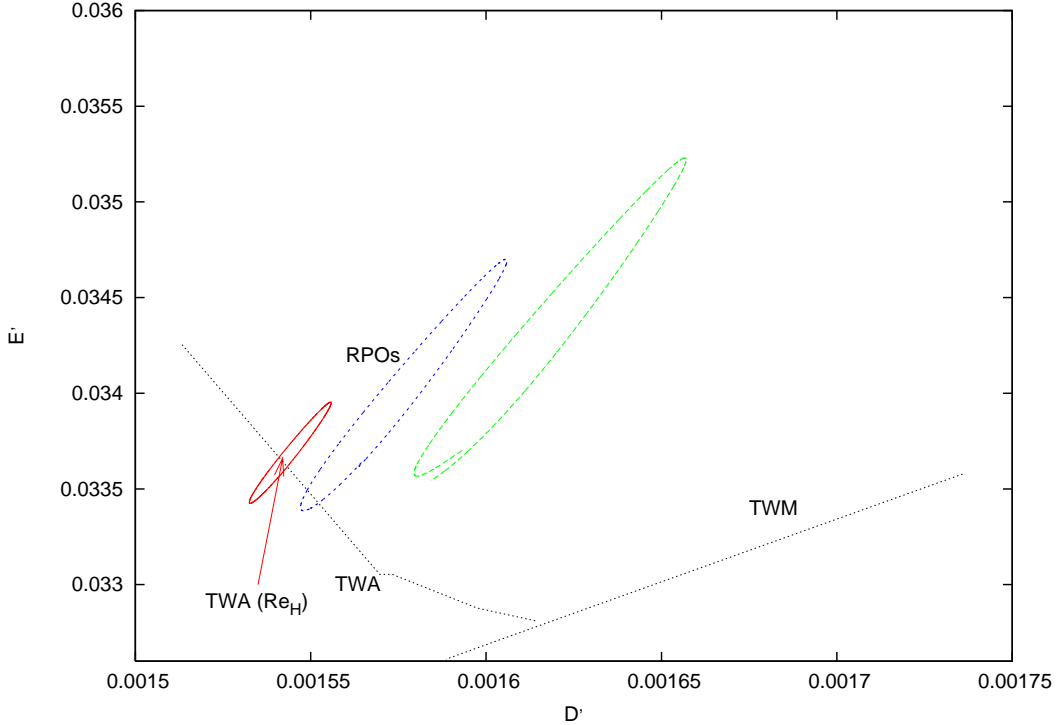}
\caption{Numerically found RPOs in the $(D',E')$ plane for different
values of $\delta=4$ (solid red) $\delta=22$ (dotted blue)
$\delta=47$ (dashed green), with $\delta=Re_{H}-Re$. The
asymmetric (TWA) and mirror-symmetric (TWM) TW branches
are also indicated (black dotted curves), parametrised by $Re$. The
TWA at $Re=Re_H$ is indicated by the red arrow. The
RPO corresponding to $\delta=47$ does not look perfectly closed,
because of numerical issues detailed in Subsection \ref{issues}.
Note that the TWA branch exists only for $Re \geq 1770$ ($\delta
\leq 15$).} \label{dissRPO}
\end{figure}

Trajectories corresponding to RPO solutions have been traced on a
two-dimensional plane $(D',E')$ for various values of $\delta$. Here
$D':=Re^{-1}\int |{\bm \nabla} \times {\bm u'}|^2 d^3x$ and
$E':=\frac{1}{2}\int |{\bm u'}|^2 d^3x$ are respectively the
dissipation and the kinetic energy of the disturbance ${\bm
u'}:={\bm u}-{\bm u}_{HP}$ to the laminar flow (see Figure
\ref{dissRPO}). Such a projection makes these RPOs look periodic,
whereas in the non-moving frame they are only relative periodic
orbits. The orbits are slightly elliptic and correspond to a slow
modulation of the TW from which they have bifurcated, whereas a TW
at a given $Re$ would appear as a dot.

\begin{figure}
\includegraphics[width=14cm]{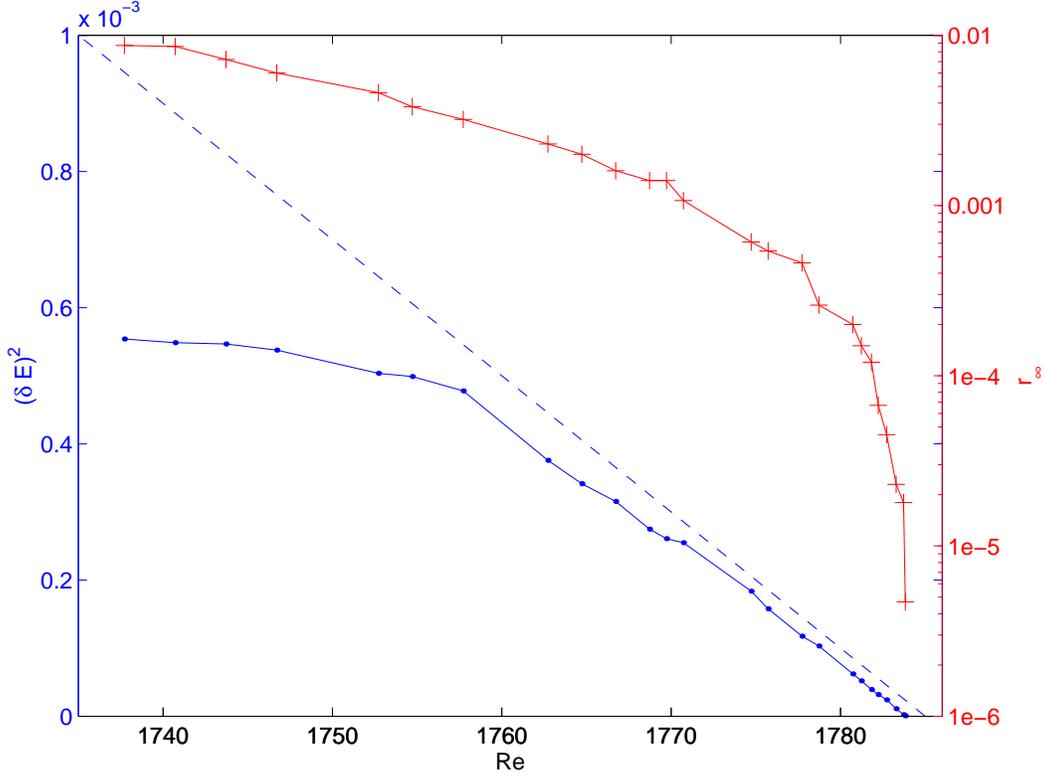}
\caption{Normalised amplitude $(\delta E)^{2}$ for the RPO as a
function of $Re$ (lower blue data), along with asymptote $(\delta
E)^2 = 2 \times 10^{-5}\delta$ where $\delta=Re_H-Re$ (blue dashed
line). The minimal normalised residual $r_{\infty}=|{\bm X}_{-\Delta
z}(T)-{\bm X}(0)|/|{\bm X}(0)|$ achieved by the Newton-Krylov
algorithm is also shown as a function of $Re$ (upper red data).
Continuation was stopped when the numerical residual reached
$10^{-2}$.} \label{deRPO}
\end{figure}

We define their normalized amplitude by :
\begin{eqnarray}
\delta E=\frac{E_{max}-E_{min}}{E_{max}+E_{min}}
\end{eqnarray}
 where the subscripts $min$ and $max$
denote extrema along a cycle of the RPO ($\delta E=0$ for a TW
solution). The further $\delta$ is away from the bifurcation point,
the larger the amplitude of the RPO, as is expected from a regular
Hopf bifurcation. The values of $T$ and $\Delta z$ both keep the
same order of magnitude. For $\delta$ up to 50, the tendency for the
orbits is to achieve excursions towards regions of higher
dissipation and energy. The energy variations along one cycle stay
small for the range of $Re$ considered. Figure \ref{deRPO} shows the
dependence of the orbit amplitude $\delta E$ on $Re$: $\delta E$
grows monotonically with the reduced parameter $\delta$ and is well
approximated by the expression $\delta E = O(\delta^{\frac{1}{2}})$
near the bifurcation point. The curve has been continued numerically
until $\delta=47$. Above this value, convergence becomes
questionable - see Figure \ref{deRPO} - as discussed in Subsection
\ref{issues} and in the conclusion.

\begin{figure}
\includegraphics[width=15cm]{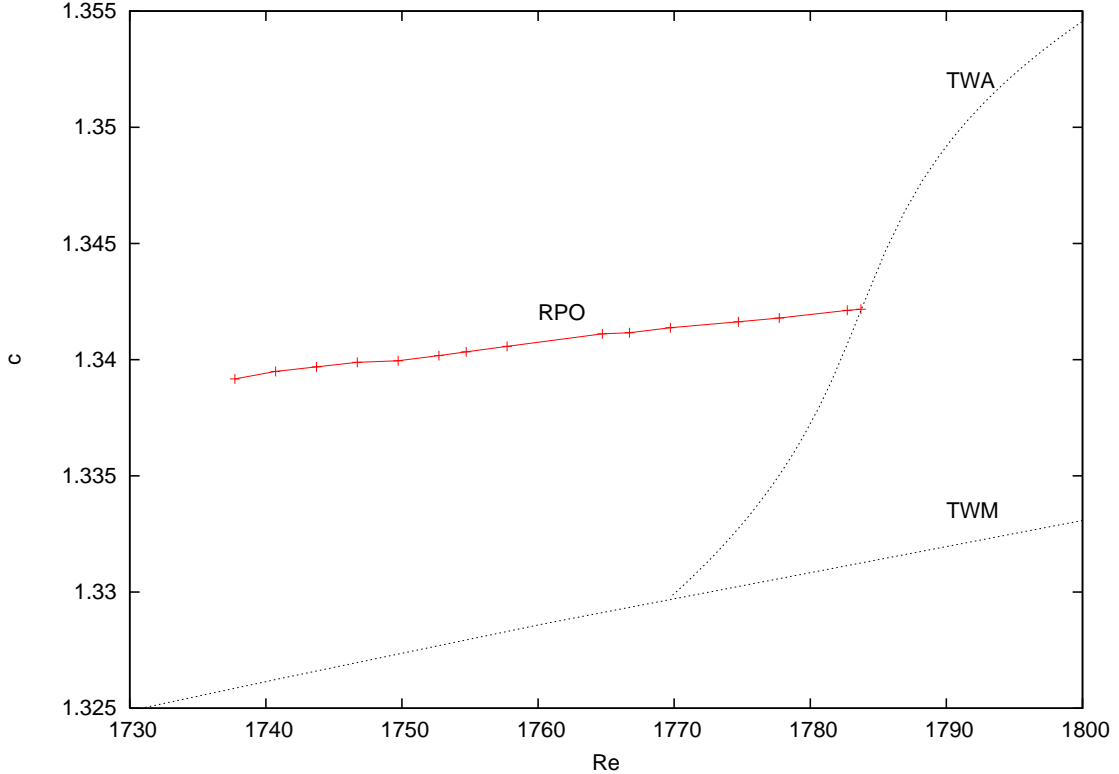}
\caption{Axial phase velocity $c$ (in units of $U$) vs. $Re$ for the
RPO (red), the asymmetric TW (TWA) and the mirror-symmetric TW (TWM)
(both in black). } \label{cRPO}
\end{figure}

The value of $\Delta z$ indicates the distance travelled by the
exact coherent structure as it propagates down the pipe in time $T$.
This defines a phase speed for the RPO,
\begin{eqnarray}
c_{RPO}:=\Delta z/T, \label{def_cRPO}
\end{eqnarray}
corresponding to the axial speed of the frame in which the orbit
would be exactly periodic. Figure \ref{cRPO} is a diagram showing
the phase velocity of both the asymmetric TW branch and the
mirror-symmetric TW branch, compared to the phase velocity of the
RPO which varies little over the range of $Re$ considered. Note that
the RPO exists at values of $Re$ below that at which the asymmetric
TW appears (see Figure \ref{cRPO}).

\begin{figure}
\includegraphics[width=8cm]{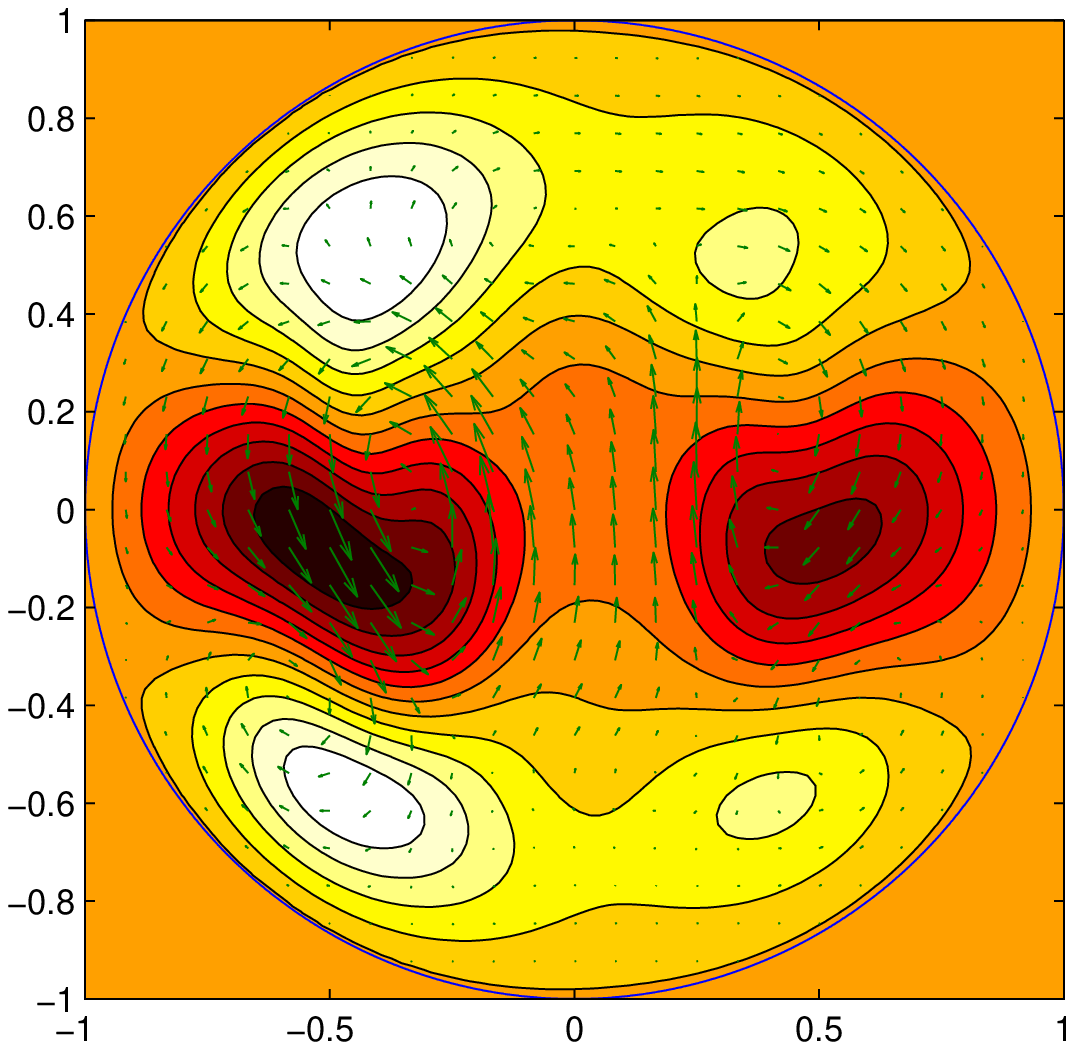}
\includegraphics[width=8cm]{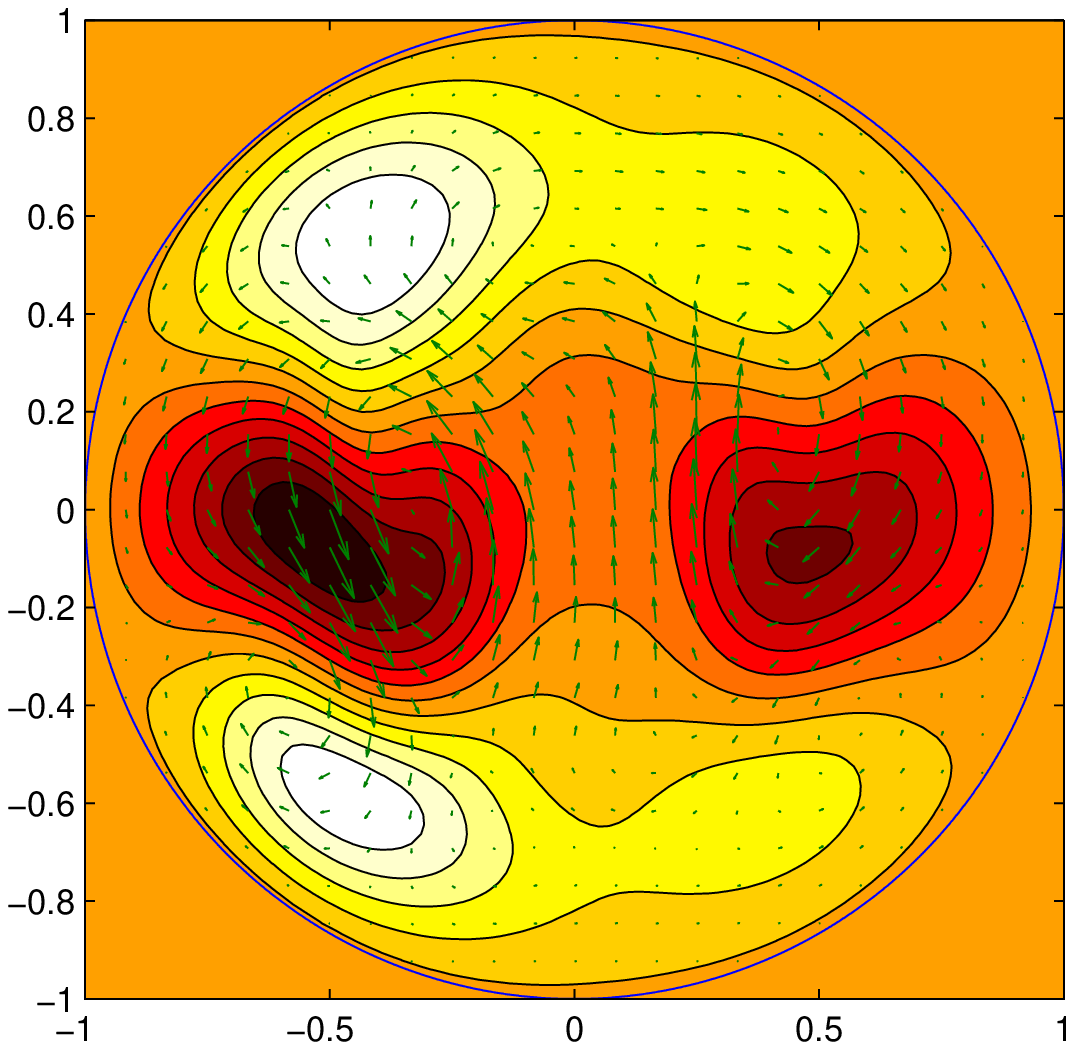}\\
\includegraphics[width=8cm]{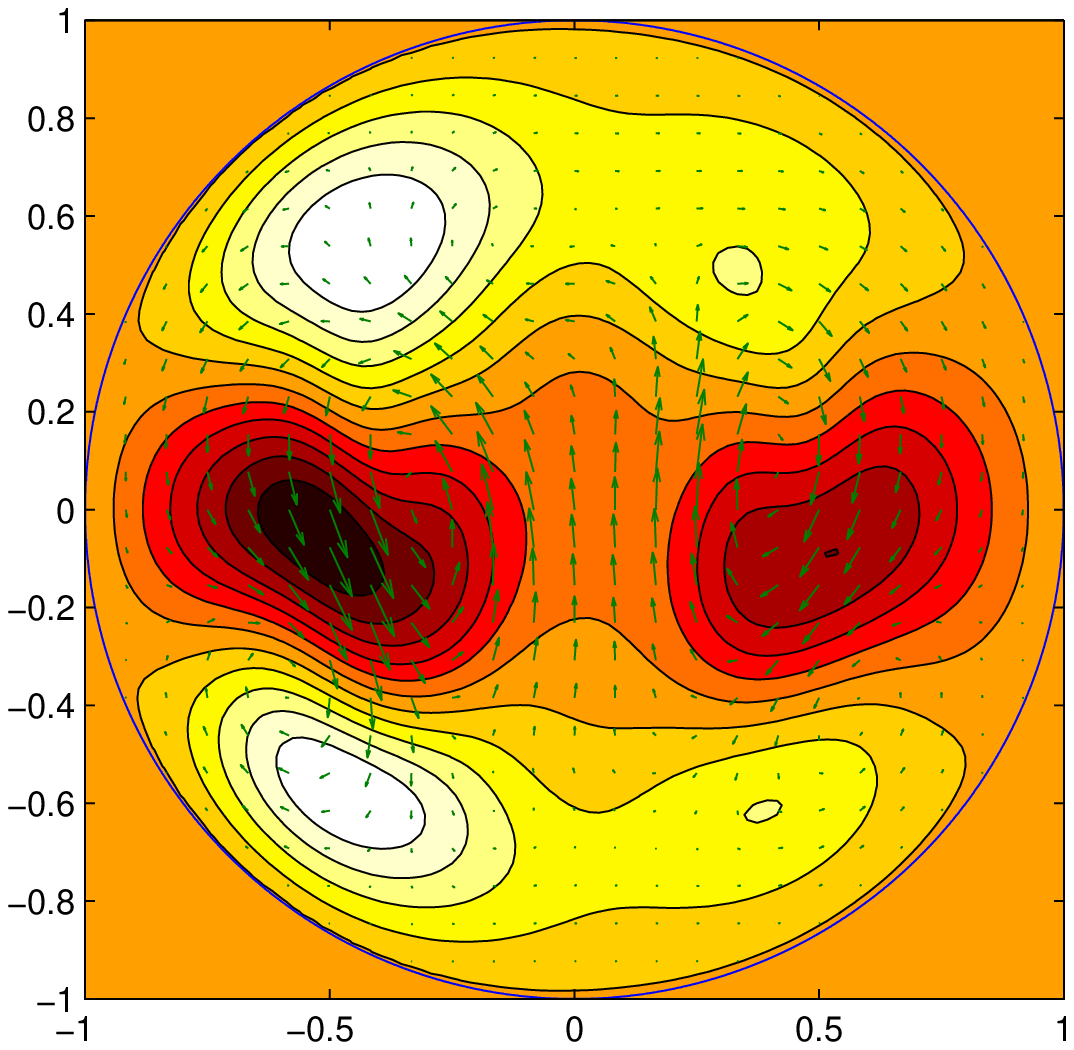}
\includegraphics[width=8cm]{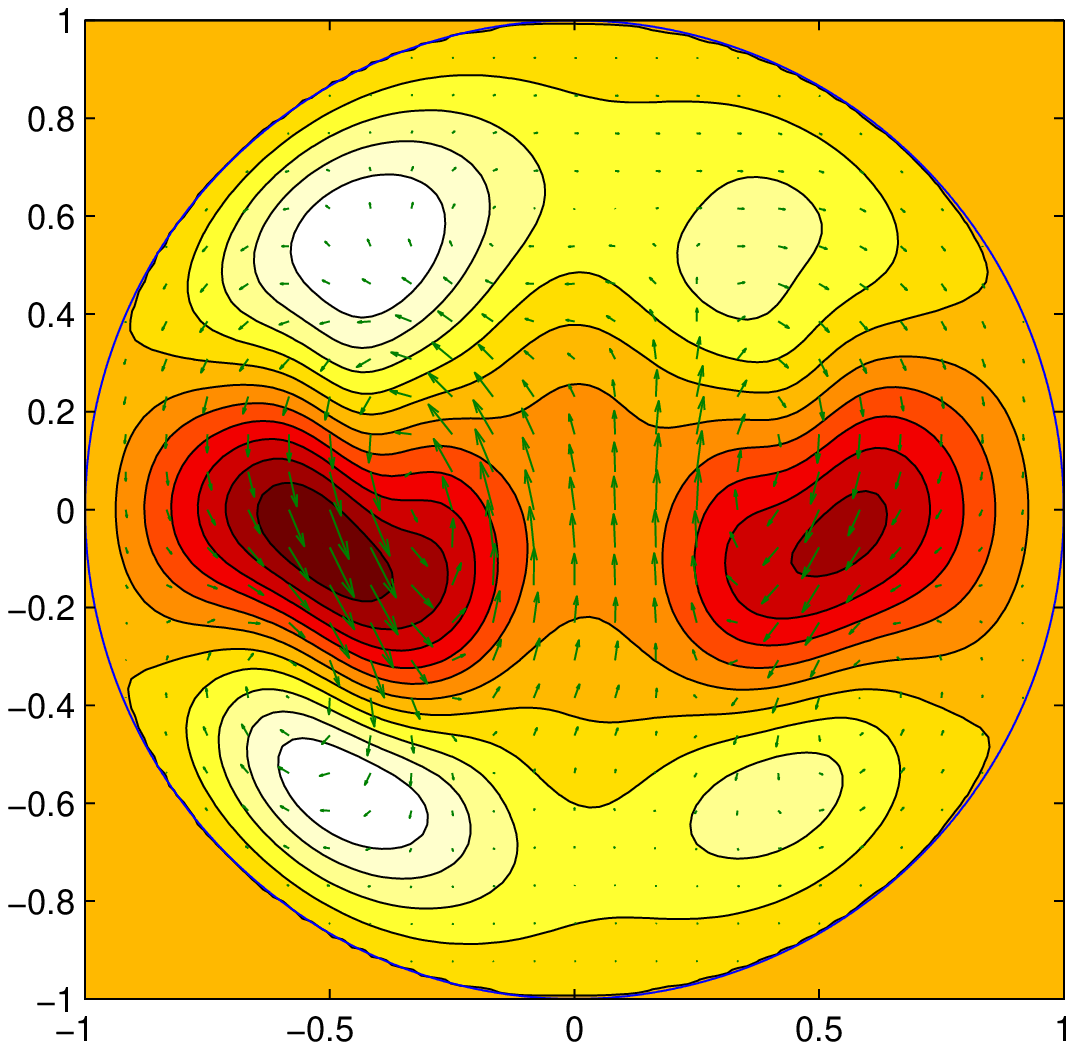}\\
\caption{Velocity field of the RPO in the moving frame, at a
cross-section defined by $z-c_{RPO}t=0$ (see Eq. \ref{def_cRPO}),
for $\delta=47$ {\bf ($Re=1738$)}. From left to right, from top to
bottom : $t=7.2$, $t=18.7$, $t=30.3$ and $t=41.8$ in units of $d/U$.
The colours and arrows are as in Figure \ref{plots0}.}
\label{plots1}
\end{figure}

Figure \ref{plots1} shows the evolution of the velocity field in a
cross-section moving with the speed $c_{RPO}$ at $\delta=47$ (i.e.
$Re \sim 1738$). This moving frame is chosen to emphasize slow-time
variations. Near the left of the cross-section, the flow is strongly
reminiscent of the pattern of the original asymmetric TW, which 
 would give 4 identical slices. There is a slight periodic
modulation of the shape and position of the low-speed streak. The
high-speed streaks close to the wall look more robust in shape
during a cycle, but their intensity fluctuates. As $\delta$
increases, the intensity inside the patches of axial velocity on the
right of Figure \ref{plots1} fluctuates more in time than the ones
of the left. The larger $\delta$, the more symmetric the
cross-sectional pattern looks (this feature is also shared with the
asymmetric TW branch as it approaches the mirror-symmetric branch).
The cross-section velocity is more steady than the axial velocity,
except in the vicinity of the weaker low-speed streak on the right
of the pipe, where the vortex looks `attached' to the streak.

\begin{figure}
\includegraphics[width=12cm]{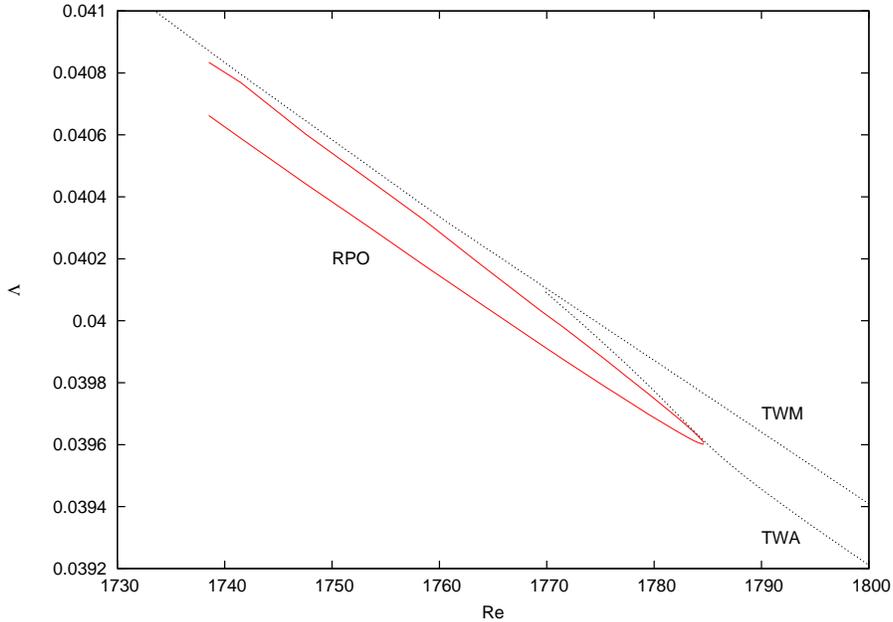}
\caption{Friction factor $\Lambda$ vs. $Re$ for the RPO (red), the
asymmetric TW and the mirror-symmetric TW (both black). In the case
of the RPO, only the extremal values $\Lambda_{min}$ and
$\Lambda_{max}$ over a cycle are shown. } \label{fricRPO}
\end{figure}

Figure \ref{fricRPO} displays the friction factor $\Lambda$, defined
using dimensional variables as
\begin{equation}
\Lambda := -\frac{dp}{dz}/(\frac{1}{2}\rho U^{2}),
\end{equation}
of the RPO branch. For any given value of $Re$, $\Lambda$ along a
cycle is always below those of the asymmetric and mirror-symmetric
TW solutions. The highest value $\Lambda_{max}$ reached during a
cycle gets closer to that of the mirror-symmetric branch as $Re$ is
decreased.

\subsection{Instability of the RPO and numerical issues \label{issues}}

While performing the continuation in $\delta$, we observed that
$r_{\infty}$ (the value around which the numerical residual
stagnates) increases steadily with $\delta$. This trend is clear
from Figure \ref{deRPO}. While animation of the velocity field does
not show any discontinuity in time for $\delta
> 25$, the projection on the $(D',E')$ plane no longer looks perfectly
closed (see Figure \ref{dissRPO}). We deliberately stopped the
continuation near $\delta \sim 50$ when
it was difficult to reduce $r_{\infty}$ below $10^{-2}$. \\

In order to understand this behaviour of $r_{\infty}$, we first need
an idea of how accurate the search for RPOs is expected to be. The
minimum residual $r_{\infty}$ is limited by both the accuracy of the
time-stepping scheme \emph{and} the natural instability of the
trajectory. Any component of any point on the numerically found RPO
is accurate to at least $O(\epsilon_{m})$, the machine precision.
After a period $T$, this numerical discrepancy has increased by a
factor $e^{\lambda T}$, with $\lambda$ being now the largest Floquet
exponent of the RPO. Indeed a useful rule-of-thumb for estimating
the lowest residual attainable is:
\begin{eqnarray}
r_{\infty} \sim O \left( n \epsilon_{m} e^{\lambda T}\right).
\label{r_infty}
\end{eqnarray}
We can illustrate this formula by simply time-stepping from a
sufficiently converged asymmetric TW at $\delta=0$. After the TW has
travelled one single pipe length, $r(t=L/c) \sim 10^{-10}$; whereas after a time $T$ (the period of the RPO), $r(T) \sim 10^{-5}$. Using expression
(\ref{r_infty}), we can estimate $\lambda_{TW} \sim \ln(10^5)/T \sim
0.26 \, U/d$. Sufficiently close to the Hopf bifurcation, both the
RPO and the TW are expected to have the same leading Floquet
exponent $\lambda$, hence the estimation
$\lambda_{RPO}(\delta=0) \sim 0.26 \, U/d$. Since the period $T$ does not vary much (less than
$1\%$) within the range of $\delta$ considered, we can extract from the data in Figure 6 a trend for the Floquet exponent of the RPO. We found the following scaling :
$\lambda_{RPO} (\delta) \sim O(\delta^{\gamma})$, with $\gamma$ very close to $\frac{1}{8}$. Unsurprisingly, the RPO is more unstable away
from the bifurcation point. A possible explanation is the stronger slow-time modulation of the
streaks (whose strong transversal velocity gradients are already
responsible for the inertial instability of the TW) as $\delta$ increases. This
harmonic modulation of the streaky field over a full cycle is
likely to increase the instability of the whole flow.\\

\section{Connection with a TW solution \label{sec:heteroc}}

A leading question is how the RPO found in this paper is related to
transition to turbulence in a pipe. The asymmetric TW is already
known to play a special role. Firstly, it lies on the
laminar-turbulent boundary (also called the `edge of chaos'
\cite{Schneider, Schneider2}), which is the set of initial
conditions separating trajectories which directly relaminarise from
those which lead to turbulent behaviour. This means that
infinitesimal perturbations to the TW can either lead to transition
or relaminarisation, depending on their exact form. Secondly, recent
studies have shown that for analogous parameters, but larger $Re$,
this asymmetric TW is generally the only exact travelling wave
solution closely visited by trajectories when constrained on the
laminar-turbulent boundary. \cite{Duguet} It is thus interesting to
see if other kind of exact recurrent states, like the RPO found here
fulfil the expectation of: (a) lying on the laminar-turbulent boundary as well;
(b) connecting to other states via heteroclinic connections;
and (c) getting transiently approached by transitional trajectories.

We have chosen to concentrate on
 $\delta=2$ ($Re=1783.6$), and perturbed the RPO by two random vectors
 in the following manner :
\begin{eqnarray}
{\bm X}(t=0) = {\bm X}_{RPO} + \epsilon \left(cos \phi \,{\bm
e}_1 + sin \phi \,{\bm e}_2 \right). \label{shoot}
\end{eqnarray}

In this expression, $\epsilon$ is a small parameter (here
$10^{-4}$), ${\bm e}_1$ and ${\bm e}_2$ are two random vectors of
norm unity, which are generic enough to possess non-zero components
along the unstable manifold of the RPO. $\phi$ is a shooting angle
which has been varied in the range $[0,2\pi]$. We notice that
several trajectories starting from the new state ${\bm X}(t=0)$
relaminarise directly while other trajectories undergo a short
`turbulent' transient, indicated by a dramatic rise in the kinetic
energy and dissipation of the disturbance to the laminar flow, and
by a loss of symmetry of the flow. Hence the RPO found for
$Re=1783.6$
is also on the laminar-turbulent boundary.\\

\begin{figure}
\includegraphics[width=12cm]{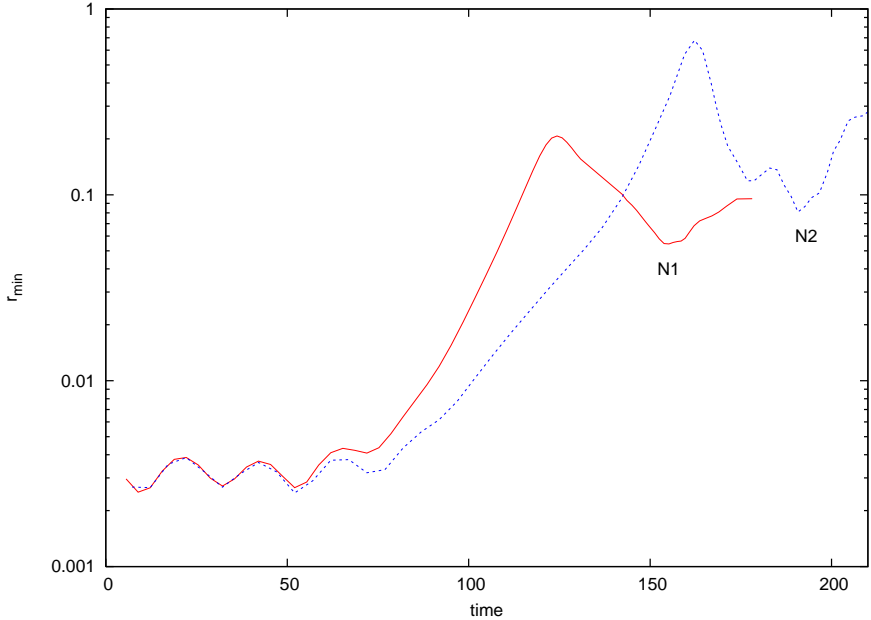}
\caption{Recurrence function $r_{min}$ vs. time for the two `edge'
trajectories $H_{1}$ (red solid) and $H_{2}$ (blue dotted), starting
from the perturbed RPO and constrained to the laminar-turbulent
boundary. The dip labelled N1 (resp.N2) near $t \sim 150$ (resp. $t
\sim 190$) indicates an approach towards a TW state. }
\label{rminRPO}
\end{figure}

Establishing heteroclinic connections between the RPO and other
exact coherent structures is an involved and technically demanding
task. We can, however, seek suggestive evidence for such links by
exploring the evolution of trajectories originating close to the RPO
and confined to remain in the laminar-turbulent boundary. This was
attempted for the RPO at $\delta=2$ by refining the angle $\phi$
defined in (\ref{shoot}) to find a phase-space trajectory which
stays near the laminar-turbulent boundary, i.e. neither directly
relaminarises nor undergoes a turbulent transient. This is done via
shooting method based on bisection of the value of $\phi$. Refining
$\phi$ up to three significant digits results in a trajectory called
$H_{1}$, which eventually relaminarises after a significantly long
transient along the laminar-turbulent boundary. For a short duration
during the transient (less than $10\,d/U$), all velocity components
oscillate with the same apparent period on a short-time scale, while
the energy reaches a plateau. Based upon previous
experience\cite{Duguet}, we recognised this as the signature of a
travelling wave solution lying nearby in phase space. The
scalar function
\begin{eqnarray}
r_{min}(t)=\min_{t'>t} \{\,
r_{t}(t'), |\frac{\partial r_{t}}{\partial t'}=0\,\}
\end{eqnarray}
where
\begin{eqnarray}
r_{t}(t'>t)=\frac{|{\bm X}(t')-{\bm X}(t)|}{|{\bm X}(t)|}.
\end{eqnarray}
was calculated along the whole trajectory. This function, closely
linked to $|{\bm G}|$, measures how recurrent a flow is at a given
location in space (ignoring the possibility of shifted recurrences).
Figure \ref{rminRPO} shows $r_{min}$ as a function of the starting
point on the edge trajectory $H_{1}$. While $t \le 60\,d/U$,
$r_{min}$ is very low (\,O($10^{-3}$)\,) and the flow is nearly
recurrent. The slow-time modulation reflects the modulation of the
flow along the RPO. Later the trajectory leaves the neighbourhood of
the RPO because of its instability, and $r_{min}$ increases up to
large values of $\approx 0.2$. At a later phase corresponding to the
plateau in energy, $r_{min}(t)$ displays a clear dip (labelled N1 in
Figure \ref{rminRPO}) down to values of $ \approx 5 \times 10^{-2}$,
and then increases again. Such low values of $r_{min}$ are never
reached if the angle $\phi$ is chosen randomly and indicate that the
edge trajectory has entered the (phase-space) vicinity of an exact
periodic solution, at which $r_{min}$ vanishes. The starting point
at $t \sim 150$ which yielded the lowest $r_{min}$ was used as an
initial state for the Newton-Krylov algorithm with $\Delta z
=2\pi/\alpha$ and $\Delta \theta =0$. This readily converged to the
asymmetric TW, shifted azimuthally by a half turn, with $r_{min}
\sim 10^{-10}$.\\

Shooting in the opposite direction, then refining properly the angle
$\phi$, leads to another `edge' trajectory $H_{2}$. The
corresponding recurrence
 function $r_{min}(t)$ is plotted in Figure \ref{rminRPO}.
 The features of $r_{min}$ are reminiscent
 of those of the trajectory $H_{1}$. A dip (labelled N2) appears near $t
 \approx 190 \, d/U$, and the corresponding state has been used
 as a starting point for the Newton-Krylov algorithm (again we look explicitly
  for a TW solution). The algorithm converged to another TW solution,
   that we call $2b\_1.5$. Close examination revealed that it is equivalent
   of the TW $2b\_1.25$ mentioned
    in Kerswell \& Tutty \cite{Tutty} and Duguet et. al. \cite{Duguet},
but with an axial wavenumber of  $\alpha=1.5$ instead of
$\alpha=1.25$. Here $\alpha=0.75=1.5/2$, so that exactly two
wavelengths of the TW fill the whole domain. This TW solution was
already mentioned in Duguet et. al. \cite{Duguet} because it is an
attractor for edge trajectories when $Re=2400$, $\alpha=1.25$, and
when the flow is constrained to a 2-fold rotationally symmetric
subspace. This attracting property is lost
 when no rotational symmetry is forced, which explains why the
 TW here first attracts and then repels the edge trajectory $H_{2}$.\\

Given the difficulty of getting the Newton-Krylov algorithm to converge
 in high dimensions unless the starting guess is sufficiently close to a solution, it is reasonable to assume that the above trajectories enter the neighbourhood of the TWs before ultimately relaminarising. Since the unstable manifold of these TWs in the laminar-turbulent boundary is of such
 small dimension, it is tempting to speculate that these visits are actually indicative of heteroclinic connections linking the RPO to the two TWs. Establishing this formally is a significant undertaking not pursued further here but the idea that trajectories link different coherent structures to produce  a saddle structure on the laminar-turbulent boundary is consistent with other recent observations.\cite{Duguet}

\begin{figure}
\includegraphics[width=16cm]{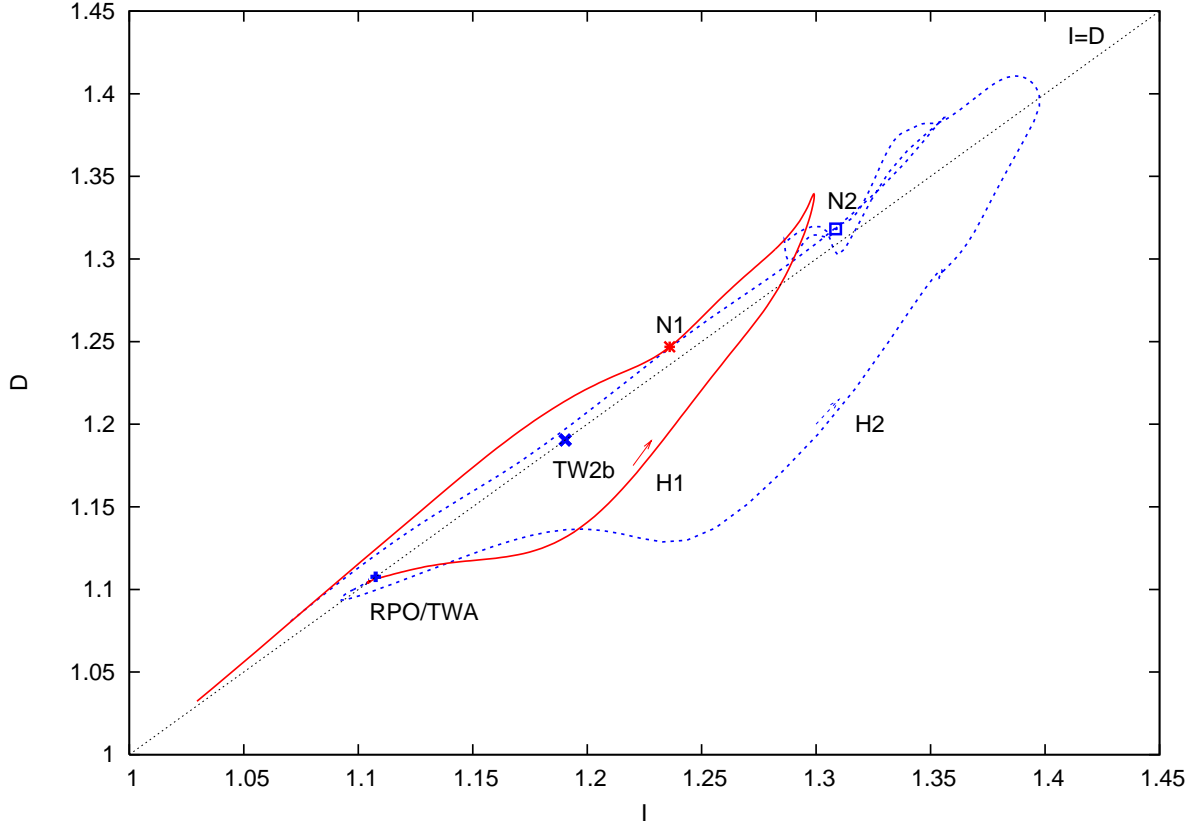}
\caption{Normalised $(I-D)$ projection of the `edge' trajectories $H_{1}$ (red) and $H_{2}$ (blue) starting from the perturbed RPO, $Re=1783.6$. Also shown are the two TW states (blue cross and dot) identified using the Newton-Krylov algorithm : TWA stands for the Asymmetric TW, TW2b for the TW solution $2b\_1.5$. The location of the RPO is undistinguishable from that of the Asymmetric TW on this plot. }
\label{heterodissRPO}
\end{figure}

The trajectories $H_{1}$ and $H_{2}$ are projected on a two-dimensional $(I,D)$
diagram in Figure \ref{heterodissRPO}. $I$ and $D$ are respectively the energy input and the total viscous dissipation in the computational domain, defined by $I:=\oint p \bm{u}\cdot\bm{n}dS$, $D:=Re^{-1}\int |{\bm \nabla} \times {\bm u}|^2 d^3x$. These two quantities are related by the energy equation 
\begin{eqnarray}
\frac{dE}{dt}=I-D,
\label{energyeqn}
\end{eqnarray}
 where $E$ is the total kinetic energy $E:=\frac{1}{2}\int |{\bm u}|^2 d^3x$ . Hence (relative) equilibria are all located on the diagonal $I=D$. $I$ and $D$ are normalised in Figure \ref{heterodissRPO} by $\bar{I}$, which is the value of $I$ for the laminar flow (hence located here at $(I,D)=(1,1)$). Both trajectories $H_1$ and $H_2$ escape from the RPO solution along its unstable
manifold, in a direction locally tangent to the laminar-turbulent
boundary. They reach much larger values of $D$ and $I$, before turning back towards the laminar state, each of them passing by the
vicinity of one of the TWs. The time of closest approach in the $(I,D)$ plane does not coincide with the minima of $r_{min}$ (e.g. $N2$ on $H2$ is not the closest point of approach to $2b\_1.5$ in the $(I,D)$ plane).  This difference, of course, highlights the ongoing
dilemma of how to select the `right' norm to measure distances in the infinite-dimensional phase space associated to the Navier-Stokes equations.

\section{Conclusion} \label{sec:disc}

We have shown that it is possible to numerically capture relative
periodic orbits (RPOs) in pipe flow when the associated dynamical system size
exceeds $10^{5}$ degrees of freedom. We have applied successfully our
method to find a RPO branch in the vicinity of the Hopf bifurcation
occurring along the asymmetric Travelling Wave branch for
$\alpha=0.75$. This exact coherent structure consists of a streaky
pattern which is modulated over one period of roughly $43\, d/U$, while
travelling approximately 60 pipe diameters downwards in the axial
direction.\\

Further evidence is presented here that the laminar-turbulent
boundary is structured around of network of exact solutions linked
to each other by heteroclinic and homoclinic connections
\cite{Tutty,Wang,Gibson,Duguet}. For $L \sim 5\,d$ ($\alpha=0.625$)
and $Re=2875$, asymmetric TW solutions are recurrently approached by
laminar-turbulent trajectories;\cite{Duguet} here for $L \sim
4.19\,d$ ($\alpha=0.75$) and $Re=1783.6$, the same conclusion seems
to hold. However, now this schematic view of edge trajectories
recurrently visiting neighbourhoods of TWs needs to be  extended to
take into account relative periodic
orbits.  All these solutions and their stable manifolds make up the laminar-turbulent boundary.\\

The main problem in accurately identifying RPOs is the exponential divergence
of a nearby orbit from the RPO. In practice, the instability of the RPO together
 with its typically long period lead to a large value for its leading characteristic
multiplier. This causes a significant loss of numerical accuracy when searching for neighbouring orbits which almost close on themselves. This effect was found to worsen
away from the bifurcation point as the RPO becomes more unstable so that it was only possible to trace the RPO branch accurately in a relatively small neighbourhood of the bifurcation point. As a result, its amplitude remains small and the RPO dynamics is close to that of the underlying TW solution. The original goal of tracing the RPO down (in $Re$) to a likely saddle node bifurcation point and then exploring the `other' branch have, unfortunately, been out of reach. Whether this `other' branch reconnects to another TW branch or becomes a more nonlinear unstable object possibly embedded in the flow's turbulent dynamics remains a challenge for the future. There are ways to improve the numerical algorithm such as a multiple-shooting technique where the orbit is subdivided into manageable pieces which now clearly need to be explored.

Physically, the instability of the RPO means that phase-space trajectories are unlikely
to spend a long time in its vicinity, despite the likelihood that the RPO is linked via  heteroclinic connections to other more attracting solutions. In particular, the probability is low that such a solution could be identified experimentally,
despite recent progress made in flow visualisation or indeed numerically as part of a direct numerical simulation. The same conclusion may well apply to all RPOs of the system.\\

At the start of this study, the stability of various lower TW
branches was considered in order to identify how frequently Hopf
bifurcations of TW branches arise. For $\alpha=0.75$, the asymmetric
TW branch undergoes only one Hopf bifurcation below $Re=5000$ which is examined in this paper. The mirror-symmetric
branch displays another one near $Re=3487$, its period near the
bifurcating point is $T=68$, which makes it even more unstable and
harder to continue than the previous one. For the same parameters,
we have checked that another known branch of TW (referred to as
$2b\_{1.25}$ \cite{Tutty,Duguet}) does not display any Hopf
bifurcations at all below $Re=3400$. Despite the fact that many more
TWs might exist, it appears that Hopf bifurcations of lower-branch
TW solutions are not frequent, and that short-period RPOs associated
with them are not generic objects. From this point of view, the fact
that mainly TW solutions have been identified during transition does
not mean that RPOs do not participate in the transitional dynamics,
but rather that their contribution is likely to be more infrequent
and fleeting. However, it might not be the case for other shear
flows such as plane Couette flow where (non-relative) periodic
orbits also bifurcate from a lower-branch steady
state.\cite{KidaNagata,Kawahara,Wang} This does not exclude either
the existence of RPO solutions entirely disconnected from the TW
solutions, which would appear via saddle-node bifurcations. Such a
(periodic) solution has been computed in plane Couette flow, and its
dynamics is embedded in the turbulent dynamics.\cite{Kida,
Viswanath} The numerical method described in Section
\ref{sec:computing} is an adequate tool to seek them providing a
good enough starting point is identified for the Newton-Krylov
algorithm.



\begin{acknowledgments}
We would like to thank A.P. Willis for sharing his time-stepping
code with us and for stimulating comments during this work. Y.D. is
supported by a Marie-Curie Intra European Fellowship (grant number
MEIF-CT-2006-024627). C.C.T.P. is supported by an EPSRC grant.
\end{acknowledgments}


\begin{references}



\bibitem{Reynolds}
O.~Reynolds.
\newblock An experimental investigation of the circumstances which determine whether the motion of water shall be direct or sinuous and of the law of resistance in parallel channels.
\newblock {\em Phil. Trans. Roy. Society}, \textbf{174}:935--982,1883

\bibitem{Mullin}
A.~Darbyshire,~T.~Mullin.
\newblock Transition to turbulence in constant mass-flux pipe flow.
\newblock {\em J. Fluid Mech. }, \textbf{289}:89--114,1995

\bibitem{Pfenniger}
\newblock W.~Pfenniger.
\newblock Transition in the inlet length of tubes at high Reynolds numbers \newblock {\em Boundary Layer and Flow Control} (ed.GV Lachman), \textbf{970}, 1961

\bibitem{Drazin}
P.G.~Drazin, W.H.~Reid.
\newblock Hydrodynamic Stability
\newblock {\em Cambridge University Press},1985

\bibitem{Waleffe1}
F.~Waleffe.
\newblock Three-dimensional coherent states in plane shear flows.
\newblock {\em Phys. Rev. Letters}, {81(19)}:4140--4143, 1998

\bibitem{Bristol}
T.~Mullin,~R.R.~Kerswell (eds).
\newblock Non-Uniqueness of solutions ot the Navier-Stokes equation and their connection with laminar-turbulent transition.
\newblock {\em Springer-Verlag},2004

\bibitem{Kerswell_nonlinearity}
R.R.~Kerswell.
\newblock Recent progress in understanding the transition to turbulence in a pipe.
\newblock {\em Nonlinearity}, \textbf{18}:17--44, 2005

\bibitem{Gibson}
J.F.~Gibson,~J.~Halcrow,~P.~Cvitanovi\'c
\newblock The geometry of state space in plane Couette flow
\newblock {\em accepted for publication in J. Fluid Mech.}, 2008

\bibitem{Hamilton}
J.~Kim,~J.~Hamilton,~F.~Waleffe.
\newblock Regeneration mechanisms of near-wall turbulence structures.
\newblock {\em J. Fluid Mech.}, \textbf{287}:317--348, 1995

\bibitem{Waleffe}
F. Waleffe
\newblock Exact Coherent Structures in Channel Flow,
\newblock {\em J. Fluid Mech.}, \textbf{435}:93-102, 2001

\bibitem{Nagata}
M.~Nagata.
\newblock Three-dimensional finite-amplitude solutions in plane Couette flow : bifurcation from infinity.
\newblock {\em J. Fluid Mech.}, \textbf{217}:519--527, 1990

\bibitem{Faisst}
H.~Faisst, B.~Eckhardt.
\newblock Traveling waves in pipe flow.
\newblock {\em Physical Review Letters}, \textbf{91}:224502, 2003

\bibitem{Wedin}
H.~Wedin, R.R.~Kerswell.
\newblock Exact coherent structures in pipe flow : travelling wave solutions.
\newblock {\em J. Fluid Mech.}, \textbf{508}:333--371, 2004

\bibitem{Hof}
B.~Hof,~C.W.H.~van~Doorne,~J.~Westerweel,~F.T.M.~Nieuwstadt,
~H.~Faisst,~B.~Eckhardt, H,~Wedin,~R.R.~Kerswell,~F.~Waleffe.
\newblock   An experimental observation of travelling waves in pipe flow
 \newblock{\em Science}, \textbf{305}, 1594-1597, 2004

\bibitem{Hof2}B.Hof, C.W.H. vanDoorne, J.Westerweel, F.T.M. Nieuwstadt.
\newblock Turbulence regeneration in pipe flow at moderate Reynolds numbers.
\newblock {\em Phys. Rev. Lett.}, \textbf{95}, 214502,2005

\bibitem{Tutty}
R.R.~Kerswell,~O.R.~Tutty.
\newblock Recurrence of travelling waves in transitional pipe flow.
\newblock {\em J. Fluid Mech.}, {584}:69--102, 2007

\bibitem{Schneider3}
T.M.~Schneider,~B.~Eckhardt,~J.~Vollmer
\newblock Statistical analysis of coherent structures in transitional pipe flow
\newblock {\em Phys. Rev. E }, \textbf{75}, 066313, 2007


\bibitem{ChaosBookRef}
F.~Christiansen,~P.~Cvitanovi\'c,~V.~Putkaradze.
\newblock Spatiotemporal chaos in terms of unstable recurrent patterns
\newblock {Nonlinearity}, {10}:55-70, 1997

\bibitem{Lopez}
V.~Lopez,~P.~Boyland,~M.T.~Heath,~R.D.~Moser.
Relative Periodic Solutions of the Complex Ginzburg-Landau Equation
\newblock {\em SIAM J. Applied Dynamical Systems}, {4};1042-1075, 2005


\bibitem{Kida}
G.~Kawahara, S.~Kida.
\newblock Periodic motion embedded in plane Couette turbulence: regeneration cycle and burst.
\newblock {\em J. Fluid Mech.}, {449}:291--300, 2001

\bibitem{Viswanath}
D.~Viswanath.
\newblock Recurrent motions wihtin plane Couette turbulence.
\newblock {\em J. Fluid Mech.}, {580}:339--358, 2007

\bibitem{Pringle}
C.C.T.~Pringle,~R.R.~Kerswell.
\newblock Asymmetric, helical and mirror-symmetric travelling waves in pipe flow.
\newblock {\em Phys. Rev. Letters}, {99}:074502, 2007

\bibitem{Duguet}
Y.~Duguet,~A.P.~Willis,~R.R.~Kerswell.
\newblock Transition in pipe flow : the saddle structure on the boundary of turbulence.
\newblock {\em accepted for publication in J. Fluid Mech.}, 2008

\bibitem{Peixinho}
 J.~Peixinho,~T.~Mullin
\newblock Decay of turbulence in pipe flow
\newblock Phys. Rev. Lett. 96, 094501 (2006)

\bibitem{Peixinho2}
J.~Peixinho,~T.~Mullin
\newblock Finite-amplitude thresholds for transition in pipe flow.
\newblock  {\em J. Fluid Mech.}, {582}:169-178, 2007

\bibitem{Willis}
A.P.~Willis,~R.R.~Kerswell.
\newblock Critical behaviour in the relaminarisation of
localised turbulence in pipe flow.
\newblock {\em Phys. Rev. Lett.}, {98}:014501, 2007

\bibitem{Willis2}
A.P.~Willis,~R.R.~Kerswell.
\newblock Coherent structures in localised and global pipe turbulence.
\newblock {\em Phys. Rev. Lett.}, {100}:0124501, 2008

\bibitem{Willis3}
A.P.~Willis,~R.R.~Kerswell.
\newblock Turbulent dynamics of pipe flow captured in a
reduced model: puff relaminarisation and localised `edge' states.
\newblock {\em resubmitted} (decision likely by proofs), 2008

\bibitem{Marques}
F.~Marqu\`es.
\newblock On boundary conditions for velocity potentials in confined flows : Application to Couette flow.
\newblock {\em Phys. of Fluids A}, {2(5)}:729-737, 1990

\bibitem{Saad}
Y.~Saad,~M.~Schultz.
\newblock GMRES: A generalized minimal residual algorithm for solving nonsymmetric linear systems.
\newblock {\em SIAM J. Sci. Stat. Comput.}, {7}:1--14, 1986

\bibitem{Walker}
C.~Eisenstat,~H.~Walker.
\newblock Choosing the forcing terms in an inexact Newton method.
\newblock {\em SIAM J. Sci. Comput.}, {17}:16--32, 1996

\bibitem{Walker2}
H.~Walker,~R.S.~Tuminaro,~J.N.~Shadid.
\newblock On backtracking failure in Newton-GMRES methods with a demonstration for the Navier-Stokes equations.
\newblock {\em J. Comp. Physics}, \textbf{180}:549--558, 2002

\bibitem{DennisSchnabel}
J.E.~Dennis,~R.B.~Schnabel.
\newblock Numerical Methods for Unconstrained Optimization and Nonlinear Equations.
\newblock {\em SIAM Classics}, 1996

\bibitem{Wang}
J.~Wang,~J.~Gibson,~F.~Waleffe
\newblock Lower branch coherent states in shear flows: Transition and control.
\newblock {\em Phys. Rev. Lett. }, \textbf{98}, 204501, 2007

\bibitem{arpack}
R.~Lehoucq,~K.~Maschhoff,~D.~Sorensen,~C.~Yang.
\newblock ARPACK Homepage
\newblock {\em http://www.caam.rice.edu/software/ARPACK/}


\bibitem{Kuznetsov}
Y. Kuznetsov.
\newblock Elements of Applied Bifurcation theory.
\newblock {\em Springer-Verlag}, 2004

\bibitem{Schneider}
B.~Eckhardt,~T.M.~Schneider,~B.~Hof,~J.~Westerweel
\newblock Turbulence Transition in Pipe Flow.
\newblock {\em Ann. Rev. of Fluid Mech.}, \textbf{39}: 447--468, 2007

\bibitem{Schneider2}
T.M.~Schneider,~B.~Eckhardt,~J.A.~Yorke
\newblock Turbulence Transition and the edge of chaos in Pipe Flow.
\newblock {\em Phys. Rev. Lett. }, \textbf{99}, 034502, 2007

\bibitem{Kawahara}
G.~Kawahara.
\newblock Laminarization of minimal plane Couette flow : Going beyond the basin of attraction of turbulence
\newblock {\em Phys. Fluids}, \textbf{17}, 041702, 2005

\bibitem{KidaNagata}
G.~Kawahara,~S.~Kida,~M.~Nagata.
\newblock Unstable Periodic Motion in plane Couette System: The Skeleton of Turbulence.
\newblock {\em IUTAM Symposium on One Hundred Years of Boundary Layer Research}, {129}:415-424, 2006

\end{references}

\end{document}